\documentclass[fleqn,usenatbib]{mnras}

\usepackage[T1]{fontenc}

\usepackage{enumitem}

\DeclareRobustCommand{\VAN}[3]{#2}
\let\VANthebibliography\thebibliography
\def\thebibliography{\DeclareRobustCommand{\VAN}[3]{##3}\VANthebibliography}

\usepackage{graphicx}	
\usepackage{amsmath}	
\usepackage{amssymb}	
\usepackage{physics}
\usepackage{natbib}
\usepackage{cleveref}
\crefformat{section}{$\S \ #2#1#3$} 
\crefformat{subsection}{$\S \ #2#1#3$}
\crefformat{subsubsection}{$\S \ #2#1#3$}
\creflabelformat{equation}{#2#1#3}

\usepackage{newtxtext,newtxmath}

\newcommand{\LCDM}{$\Lambda$CDM}

\newcommand{\densm}{\overline{\rho}}

\newcommand{\oM}{\Omega_{m, 0}}

\newcommand{\ob}{\Omega_{b, 0}}
\newcommand{\onu}{\Omega_{\nu, 0}}
\newcommand{\cosmosis}{\texttt{\textsc{CosmoSIS}}}
\newcommand{\multinest}{\texttt{\textsc{MultiNest}}}
\newcommand{\neff}{n_{\mathrm{eff}}}
\newcommand{\sotto}{\sigma_8}
\newcommand{\lymana}{Lyman-$\alpha$}

\newcommand{\skm}{\mathrm{s}\,\mathrm{km}^{-1}}
\newcommand{\F}{\langle F \rangle}

\defcitealias{Vid2017}{I17}
\defcitealias{Bocquet_2019}{B19}
\newcommand{\Vid}{\citetalias{Vid2017}}
\newcommand{\Boc}{\citetalias{Bocquet_2019}}

\title[Weighing Cosmic Structures with GCs and the IGM]{Weighing Cosmic Structures with Clusters of Galaxies and the Intergalactic Medium}

\author[M. Esposito et al.]{
Matteo Esposito,$^{1,2}$\thanks{esposito@mpe.mpg.de}
Vid Ir\v{s}i\v{c},$^{3,4}$
Matteo Costanzi,$^{2,5,6}$
Stefano Borgani,$^{2,5,6,7}$
Alexandro Saro,$^{2,5,6,7}$ 
\\~\\
{\rm {\LARGE
Matteo Viel$^{5,6,7,8}$}}
\\~\\
$^{1}$MPE - Max-Planck-Institut für extraterrestrische Physik,Postfach 1312, Giessenbachstr., 85741 Garching, Germany\\
$^{2}$Astronomy Unit, Department of Physics, University of Trieste, via
Tiepolo 11, I-34131 Trieste, Italy\\
$^{3}$Kavli Institute for Cosmology, University of Cambridge, Madingley Road, Cambridge CB3 0HA, UK\\
$^{4}$Cavendish Laboratory, University of Cambridge, 19 J. J. Thomson Ave., Cambridge CB3 0HE, UK\\
$^{5}$ IFPU - Institute for Fundamental Physics of the Universe, Via Beirut 2, 34014 Trieste, Italy\\ 
$^{6}$ INAF - Osservatorio Astronomico di Trieste, via G. B. Tiepolo 11, I-34143 Trieste, Italy\\
$^{7}$ INFN - Sezione di Trieste, Trieste,  Italy\\
$^{8}$ SISSA, Via Bonomea 265, 34136 Trieste, Italy\\
 }

\date{Accepted XXX. Received YYY; in original form ZZZ}

\pubyear{2022}

\begin{document}
\label{firstpage}
\pagerange{\pageref{firstpage}--\pageref{lastpage}}
\maketitle

\begin{abstract}
We present an analysis aimed at combining cosmological constraints from number counts of galaxy clusters identified through the Sunyaev-Zeldovich effect, obtained with the South Pole Telescope (SPT), and from Lyman-$\alpha$ spectra obtained with the MIKE/HIRES and X-shooter spectrographs. The SPT cluster analysis relies on mass calibration based on weak lensing measurements, while the Lyman-$\alpha$ analysis is built over mock spectra extracted from hydrodynamical simulations.
The resulting constraints exhibit a tension ($\sim 3.3\sigma$) between the low $\sigma_8$ values preferred by the low-redshift cluster data, $\sigma_8=0.74 ^{+0.03}_{-0.04}$, and the higher one preferred by the high-redshift Lyman-$\alpha$ data, $\sigma_8=0.91 ^{+0.03}_{-0.03}$. We present a detailed analysis to understand the origin of this tension and to establish whether it arises from systematic uncertainties related to the assumptions underlying the analyses of cluster counts and/or Lyman-$\alpha$ forest. 
We found this tension to be robust with respect to the choice of modelling of the IGM, even when including possible systematics from unaccounted sub-Damped Lyman-$\alpha$ (DLA) and Lyman-limit systems (LLS) in the Lyman-$\alpha$ data. We conclude that to solve this tension would require a large bias on the cluster mass estimate, or large unaccounted errors on the Lyman-$\alpha$ mean fluxes.
Our results have important implications for future analyses based on cluster number counts from future large photometric surveys (e.g. Euclid and LSST) and on larger samples of high-redshift quasar spectra (e.g. DESI and WEAVE surveys). If confirmed at the much higher statistical significance reachable by such surveys, this tension could represent a significant challenge for the standard $\Lambda$CDM paradigm.
\end{abstract}




\section{Introduction}
In the last decades, a variety of probes of the large-scale structure (LSS) of the Universe have been widely used to study fundamental cosmology, reaching now competitive constraints with respect to other classical cosmological probes like the Cosmic Microwave Background (CMB) \citep[and references therein]{DESY3, eBOSS2021, KiDS}.
Among all the LSS probes, galaxy clusters and the intergalactic medium (IGM) are of particular interest in the context of cosmology, given their complementarity in terms of both redshift range and overdensity range probed. Indeed, at $z \lesssim 1.5$ galaxy clusters mark the largest cosmic perturbations that evolved in the non-linear regime and reached virial equilibrium. On the other hand, the Lyman-$\alpha$ forest traces at $2\lesssim z\lesssim 6$ structures over a wide range of scales which are still in the quasi-linear or mildly non-linear regime. 

More precisely, 
the galaxy clusters
number density evolution can be used to infer cosmological models by comparing the expected number of clusters of a certain mass and at a certain time obtained with semi-analytical methods, with the observed counts.
Together with their detection, the mass determination of the clusters of galaxies is then of paramount importance for cluster cosmology. However, the techniques used to "weigh" these clusters individually
are either observationally very expensive or not very accurate. Especially when dealing with large cluster surveys, it is not possible to apply the former on each of the detected objects and thus one often relies on the scaling relations that link the observable on which the survey is based to the cluster mass. Since the cluster mass calibration is the main limitation in current surveys, it is then of fundamental importance to understand the possible biases and uncertainties of such mass-observable relations as they can alter the cosmological parameter constraints. For this reason a lot of effort has been put into their calibration and refinement in the last years (see e.g. \citealt{Allen11,Kravtsov&Borgani} for reviews on cluster cosmology).
Among the different observables used to construct cluster surveys, the thermal Sunyaev-Zel'dovich effect (SZ, \citealt{SZ})\footnote{The SZ signal arises from the scatter of the CMB photons off the hot electrons of the intracluster medium (ICM)}, has a twofold advantage: firstly, the SZ signal is independent of cluster redshifts; secondly, it is proportional to the (line-of-sight integrated) pressure of the intra-cluster plasma, a quantity closely linked to cluster mass through hydrostatic equilibrium. For this reason, in the last years many sub-millimiter-wave surveys (ACT, SPT, Planck; see e.g. \citealt{Bleem2015,PlanckSZ,Hilton_18}) have been used to construct cluster catalogs to be used to constrain cosmological parameters.

Similarly, the IGM has been widely studied in the last decades thanks to the development of ever improving semi-analytical and numerical models. Its main manifestation, the \lymana{} forest, is a feature in the spectra of high redshift quasars consisting of a series of absorption lines, which can be used to track the distribution of neutral hydrogen along the line of sight. Hydrodynamical simulations are nowadays accurate enough to produce reliable mock spectra of the \lymana{} forest for cosmological studies. However, the observational properties of the \lymana{} forest also depend on the IGM thermal history and thus this must also be varied, along with cosmological parameters, in order to explore all the possible scenarios. More precisely, the effects of varying the thermal state of the IGM can mimic, at small scales, that induced by variations in the cosmological parameters; the resulting degeneracy between cosmology and IGM thermal history leads to looser constraints on cosmological parameters. In the last years, large surveys of low-resolution and low signal-to-noise quasars spectra (SDSS, BOSS, see e.g. \citealt{McDonald_2005,BOSS_2013,Chabanier_19,eBOSS_2020}) as well as moderate size samples with tens of high signal-to-noise and high-resolution spectra (VLT, HIRES/KECK, see e.g. \citealt{Viel2004b,Viel13,Vid2017,Boera19, Walther19,Karacayli21}) have been used to constrain the structure formation cosmological parameters on scales of 0.1 - 10 Mpc/h (see e.g. \citealt{review_McQuinn} for a review on IGM).

Combining the \lymana{} spectra with the cluster number counts can help improving the parameter inference by breaking the degeneracies between different parameters in each data set (see \citealt{Costanzi_2014} for a previous attempt at combining these two probes). As mentioned, we expect this particular combination to be effective in view of the complementarity of clusters and \lymana{} forest both in redshift ranges and dynamical regimes covered: while clusters of galaxies are highly dense (typically about 200 times denser than the average density of the Universe, up to their outer radius) and large structures observed at low or intermediate redshifts ($z \sim 0 - 2$), the quasars that show a \lymana{} forest can be found at much higher redshifts ($z \sim 2 - 5$) and the hydrogen that is responsible for the absorption lines trace much lower densities ($\lesssim 10$ times denser than the average density of the Universe). 
Another interesting sinergy between clusters and IGM is provided by investigations in which the thermal evolution of the cosmic web is followed down to small redshifts allowing for some pre-heating of the IGM, as in \cite{borgani09}.

This paper aims at exploiting such complementarity and in particular at investigating the compatibility of two recent analyses that use galaxy clusters and the Lyman-$\alpha$ flux power spectrum for cosmological parameter inference: the work of \cite{Bocquet_2019} which uses number counts of clusters detected by the South Pole Telescope (SPT) through the Sunyaev-Zel'dovich (SZ) effect, and that of \cite{Vid2017} which studies \lymana{} forest spectra obtained with the Very Large Telescope (VLT) and the Keck and Magellan telescopes. These two analyses rely on high quality data as well as sophisticated state-of-the-art analytical and numerical methods. Therefore, 
their comparison acts as a test  of the predictions of the standard $\Lambda$CDM model on cosmic growth over a wide redshift range, while their combination has the potential to provide tight constraints on cosmological parameters. On the other hand, a possible tension between such two probes could point toward the existence of unveiled systematics in the analysis methods or even the need of extensions of the standard cosmological model.

Both the \lymana{} spectra and the cluster number counts can constrain the late-time amplitude of the matter power spectrum, here parametrized with $\sigma_8$, with past works having shown that the former typically prefers somewhat higher values for this parameter - with respect e.g. to the Planck results \citep{Planck18-VI} - while the latter usually prefer lower ones. The purpose of the analysis presented in this paper is in fact to carry out an updated analysis of recent data on cluster number counts and \lymana{} data to assess whether such a tension persists and, in that case, quantify its significance and attempt to trace its origin.

The structure of the paper can be summarized as follows. We describe in \cref{sec:data} the observational data sets used for cluster number counts and \lymana{} data, and the numerical simulations used for the cosmological exploitation of the latter. In \cref{sec:methods} we present the methods of analysis and discuss the resulting cosmological constraints in \cref{sec:results}. Finally, we comment such results and provide our conclusions in \cref{sec:conclusions}.

\section{Data and simulations} \label{sec:data}
\subsection{\texorpdfstring{\lymana{}}{lyman} forest: XQ-100, MIKE, HIRES} \label{subsec:data_lyman}
The \lymana{} analysis relies on two different and complementary data sets: the XQ-100 sample \citep{XQ100_data, XQ100_analysis}, which provides 100 medium resolution (R $\simeq 5000$) and signal-to-noise (S/N $\simeq 30$) spectra and the HIRES/MIKE sample \citep{Viel13}, consisting of 25 high resolution (R $\simeq 80000$) and high signal-to-noise (S/N $>80$) spectra.

\paragraph*{XQ-100 sample:}
We used 100 QSO spectra from the XQ-100 Legacy Survey, observed with the X-Shooter spectrograph on the Very Large Telescope (VLT). These are medium resolution and signal-to-noise spectra with emission redshifts $3.5 < z_{\rm em} < 4.5$. 
The main quantity extracted from these data is the flux power spectrum, $P_F$, i.e. the power spectrum of the overdensity of the normalized flux of the quasar, along the line of sight (see e.g. \citealt{review_McQuinn}). Its extraction from the QSO spectra is modeled with the help of mock data from hydrodynamic simulations, allowing an accurate estimation of the flux power at $z=3, 3.2, 3.4, 3.6, 3.8, 4, 4.2$ for 19 linearly separated $k$-bins in the range $0.003-0.057$ s\,km$^{-1}$. 
The value of the $P_F$ is thus available for a total of 133 points in the $(k, z)$ space.
The errors and correlations between the values of the $P_F$ are summarized in the covariance matrix, which is obtained through the use of the bootstrap method. 
However, in order to correct for the underestimation of the variance due to the limited number of sightlines used (see e.g. \citealt{Rollinde13, Ir_2013,Vid2017,Wilson21}), 
we multiplied the full matrix by a factor of 1.3, which has been obtained by a comparison of different methods for estimation of the covariance matrix. We refer to \cite{XQ100_data} for further details on the data and to \cite{XQ100_analysis} for the flux power spectrum extraction.

\paragraph*{HIRES/MIKE sample:}
This sample consists of 25 high-resolution QSO spectra with emission redshifts $4.5 < z_{\rm em} < 6.5$. For 14 of the observed objects, the spectra were taken with the Keck High Resolution Echelle Spectrometer (HIRES), while the remaining 11 were taken with the Magellan Inamori Kyocera Echelle (MIKE) spectrograph on the Magellan Clay telescope. For both data sets, the measured $P_F$ is available at $z=4.2, 4.6, 5.0, 5.4$\footnote{The $P_F(k, z)$ at $z=5.4$ is only available for the HIRES sample as the corresponding redshift bin in the MIKE data set has a very low number of sightlines.} for 7 linearly separated log$k$-bins in the range $0.005 - 0.079$ s\,km$^{-1}$ resulting in a total of 49 data points. Following \citet{Viel13}, similarly to what we did for the XQ-100 data set, we used a correction factor of 1.5 for the covariance matrix of the flux power spectrum. A detailed description of this data set can be found in \cite{Viel13}.


\paragraph*{Simulations:} \label{sec:simulations}
Along with observational data, a set of simulations is used in the analysis (based on Sherwood simulations - \citealt{Sherwood}; details in \citealt{Vid2017}). 
These simulations were performed with the GADGET-3 code, an updated version of the publicly available GADGET-2 code \citep{GADGET}. Besides radiative cooling and the effect of a redshift dependent UV background, a simple star formation criteria is used, converting gas particles above an overdensity 1000 and with a temperature below $10^5$ K into stars (i.e. collisionless particles). 

The reference model simulation is a box with $2\times768^3$ gas and dark matter particles and a box size of $20/h$ comoving Mpc, in a flat \LCDM{} universe with $\oM=0.301$, $\ob=0.0457$, $n_s=0.961$, $H_0=70.2\mathrm{km}\, \mathrm{s}^{-1}$ and $\sigma_8=0.829$. 
When using the flux power spectrum to constrain cosmological parameters, one can exploit the fact that they impact on it through the amplitude and slope of the matter power spectrum, $P_m(k)$ at the scales probed by the \lymana{} forest \citep{McDonald_2005,Vid2017,Pedersen20}. Therefore, we chose 
to use the amplitude of the late time matter power spectrum, $\sigma_8$, and its effective slope, $\neff = \mathrm{d}\ln{P_m(k)}/\mathrm{d}\ln{k}$, evaluated at $k = 0.005$ $\skm$, as the parameters that describe our cosmological models.
In order to sample different cosmologies, simulations have been run for five different values of both $\sigma_8=0.754, 0.804, 0.829, 0.854, 0.904$ and $\neff=-2.3474, -2.3274, -2.3074, -2.2874, -2.2674$.

As for the thermal history of the IGM, its effect is treated by modifying the background photo-heating rates to tune the simulations to different values for the temperature at mean density $T_0(z=3.6)=7200, 11000, 14800 \mathrm{K}$ and the slope $\gamma(z=3.6)=1.0, 1.3, 1.5$ of the temperature-density relation of the low density IGM: $T=T_0(1+\delta)^{\gamma-1}$. These calculations are based on the \cite{Haardt2012} model for a ionising uniform UV background.
The cooling model takes into account various processes as collisional excitation cooling, bremsstrahlung cooling and inverse Compton cooling off CMB photons, but do not include line cooling as these are expected to have low impact on the \lymana{} forest spectra \citep{Viel13b}. For the same reason, galactic winds and AGN feedback are also not considered in the simulations.

To account for possible spatial fluctuations in the UV ionising background at high redshift, an inhomogeneous flux component arising from a rare QSO population is inserted into the simulations. This is obtained by identifying halos with a friend-of-friend algorithm and assigning them a luminosity given by the QSO luminosity function. Then a ionising flux bubble is computed from each of the identified halos. See \cite{f_QSO} and the appendix in \cite{Viel13} for details on the method. In our analysis the effect of UV fluctuations is parametrized in terms of $f_{QSO}=0, 0.5, 1$, defined as the fraction of volume averaged hydrogen photo-ionisation rate arising from a fluctuating QSO component.

For each simulation, we also varied the mean transmitted flux $\F$, by rescaling the optical depth ($\tau_{\mathrm{eff}} = -\ln{}\F$) to the observed values with some bias: $\tau_{\mathrm{eff}} = f\cdot\tau_{\mathrm{obs, eff}}$ with $f \in [0.3, 1.5]$ and $\tau_{\mathrm{obs, eff}}$ being the measurements of \cite{BOSS_2013}. Note that the variation of the mean flux is a post-processing procedure aimed at capturing the uncertainty in the photo-ionization rate, and does not require other simulations.

Finally, the redshift of reionization is also varied in the simulations with $z_{\rm rei}=7, 9, 15$. This is not well constrained by our data, but has an impact on the thermal history of the IGM, so it is important to marginalize over it.

A summary of the parameters varied in the simulations is shown in \cref{tab:sim_params}. Each parameter has been varied once at a time with the exception of the thermal state parameters, $T_0$ and $\gamma$ which are varied in a 3x3 grid to include all the possible 9 combinations. This in turn gives 1 reference model, 4 models varying $\sotto$, 4 models varying $\neff$, 8 models varying the thermal state parameters, 2 models varying $z_{rei}$ and 2 models varying $f_{QSO}$, for a total of 21 models.

\begin{table}
    \centering
    \begin{tabular}{c c c}
    \hline
    \hline
        Parameter & Values\\
    \hline
        $\sotto$ & [0.754, 0.804, 0.829, 0.854, 0.904] \\
        $\neff$ & [-2.3474, -2.3274, -2.3074, -2.2874, -2.2674] \\
        $T_0 (z=3.6)\;\mathrm{[K]}$ & [7200, 11000, 14800] \\
        $\gamma (z=3.6)$ & [1.0, 1.3, 1.5] \\
        $f_{QSO}$ & [0, 0.5, 1.0] \\
        $z_{\rm rei}$ & [7, 9, 15] \\
    \hline
    \hline
    \end{tabular}
    \caption{Summary of the parameters varied in the simulation suite used for the \lymana{} analysis. The mean flux parameters are not listed here because they are varied in the post-processing when extracting the mock spectra and not in the simulations. These are varied by a multiplicative factor of [0.3 - 1.5] around the reference model of \protect\cite{BOSS_2013}}
    \label{tab:sim_params}
\end{table}

\subsection{Galaxy clusters: SPT-SZ, WL masses} \label{subsec:data_clusters}
The cluster sample is made of 343 clusters identified through the Sunyaev-Zeldovich effect from the 2500 deg$^2$ SPT-SZ survey, which have been optically confirmed and have redshift measurements \citep{Bleem2015}. Mass estimates rely 
on weak lensing shear profiles obtained for 19 clusters by the Magellan/Megacam imager and for 13 clusters by the Hubble Space Telescope \citep{Megacam_sample, HST_sample}. 


\paragraph*{The SZ selected cluster sample:}
The SPT-SZ survey is provided by the South Pole Telescope (SPT) observing the sky in three millimeter wavebands centered at 95, 150 and 220 GHz. 
It has a field of view of $\sim 1$ degree and a resolution of $\sim 1$ arcmin, which allows deeper observations than those obtained with satellites used for all-sky maps (WMAP, Planck); this makes it well suited for the detection of high-mass clusters through the SZ-effect from $z\geq0.2$ up to the highest resdhifts at which clusters exist. The survey spans a contiguous 2500 deg$^2$ area within the boundaries $20\mathrm{h}\leq \mathrm{R.A.}\leq7\mathrm{h}$ and $-65^{\circ}\leq\mathrm{Dec.}\leq -40^{\circ}$.

The cluster extraction procedure (see \citealt{Williamson11} for details) uses a map-matching filter to discriminate the cluster SZ signal from spurious effects due to atmospheric/instrumental and astrophysical noise. 
The maps were filtered at 12 different cluster scales, varying the reference radii of the adopted spherical model; following the approach of Vanderlinde et al. \citeyearpar{Vanderlinde10} the detection significance $\xi$, used as the primary SZ observable, is defined as the highest signal-to-noise value associated with a decrement in the SZ map, across all filter scales.

The cluster sample used in this work consists of 343 clusters extracted from the full SPT-SZ catalogue presented in \cite{Bleem2015} for which redshift measurements are available and restricted to $z>0.25$ and $\xi>5$, achieving an expected and measured purity of 95 per cent.



\paragraph*{WL mass estimates:}
We exploit in the analysis WL shear measurements for 32 clusters: 
19 of these were observed with the Megacam imager mounted on the Magellan Clay telescope, spanning the redshift range $0.29 \leq z \leq 0.69$ (\citealt{Megacam_sample}), while the remaining 13 at redshifts $0.576 \leq z \leq 1.132$ were observed with the Advanced Camera for Surveys on the Hubble Space Telescope (\citealt{HST_sample}). The data used in our analysis are the reduced tangential shear profiles
and the estimated redshift distribution of the source galaxies. 
%
Details on the data reduction and analysis can be found in the aforementioned works.


\section{Analysis and methods} \label{sec:methods}
\subsection{Cosmological parameters} \label{subsec:cosmo_params}
Our analysis is performed through \cosmosis{} \citep{cosmosis}, a code for cosmological parameter estimation that allows one to combine different data sets, taking advantage of its modularity.
In order to compare the cosmological constraints to be obtained from the \lymana{} analysis to those to be derived from the SPT cluster number counts, for both analyses we sampled the full $\nu \Lambda CDM$ cosmological parameter space: ($A_s, n_s, \oM, \Omega_{b, 0}h^2, \Omega_{\nu, 0}h^2, h, \tau$).
Here $A_s$ is the amplitude of the linear matter power spectrum, $n_s$ the primordial spectral index, $\oM, \Omega_{b,0}$ and $\Omega_{\nu,0}$ the values at the present time for the density parameters associate to the total matter content, baryons and massive neutrinos, respectively, while $h$ is the Hubble constant in units of 100 km s$^{-1}$Mpc$^{-1}$. 
Based on these parameters, at each step in both analyses we included a CAMB \citep{CAMB} \cosmosis{} module, to retrieve the linear matter power spectrum and use it for the evaluation of the expected cluster number density and for calculating its amplitude ($\sotto$) and effective slope ($\neff$) which enter the \lymana{} likelihood.

\subsection{\texorpdfstring{\lymana{}}{lyman} forest}  \label{subsec:methods_Lya}
In order to perform our analysis with the help of \cosmosis{}, we wrote a \cosmosis{} module based on the same data and likelihood presented in \cite{Vid2017} (henceforth \citetalias{Vid2017}).
As a first check, we verified that results obtained using our \cosmosis{} module are fully consistent with those originally obtained by \citetalias{Vid2017}. A short presentation of such comparison can be found in \cref{apx:check_lyman}.
%

The cornerstone of the pipeline is the likelihood evaluation for the simulated model. The likelihood function we used is a simple Gaussian multi-variate, which is computed by comparing the simulated flux power spectra, $P_F(k, z)$, for a given set of parameters, with the observed ones. The expected values are obtained through interpolation between points of the available simulations grid in the parameter space (see \cref{subsec:data_lyman}). According to \Vid{} the interpolation error is small ($<$ 5\%) and does not contribute much to the total error budget of the current data. In fact, it does not bias the results and is thus ignored in our analysis. It is however important to notice that this upper bound on the interpolation error has been tested only in the explored parameter space. This may not be the case if extrapolation out of the parameter space occurs (see the comment on the constraint on $\neff$ in \cref{subsec:Lya results}).

The cosmological simulations employed are labelled by three redshift-dependent parameters ($\gamma, T_0, \F$, see \cref{sec:simulations}) which can be modelled, in our analysis pipeline, as power-laws of $(1+z)$ or by assigning to each of them independent values within each redshift, eventually setting a maximum difference between the values taken at adjacent bins. In all the analyses presented here, unless otherwise specified, we used independent values within each of the 10 redshift bins for the mean flux $\F$, while we assume power-laws of $(1+z)$ for the amplitude $T_0$ and the slope $\gamma$ of the IGM equation of state: 
\begin{align}
     T_0 (z) & = T_0^A [(1+z)/(1+z_p)] ^{T_0^S}; \nonumber \\
    \gamma (z) & = \gamma^A [(1+z)/(1+z_p)] ^{\gamma^S}\,,
\end{align}
where the superscripts $A$ and $S$ refer to the amplitude and slope of the power-law relation and $z_p$ is the redshift pivot, here set to $z_p = 4.2$. 

Given its impact on the IGM thermal history, in our analysis we also varied the redshift of reionization $z_{\rm rei}$. However, this parameter does depend on one of the employed cosmological parameters: the optical depth at recombination $\tau$. Moreover, $z_{\rm rei}$ is also weekly correlated with the temperature at mean density of the IGM, $T_0$. Hence, we modelled these two dependencies in order to reduce the sampled parameter space and map them one to another. Looking at our simulations, we have found that linearly interpolating the $z_{\rm rei}-\tau$ and $z_{\rm rei}-T_0$ relations independently, results in these parametrization:
\begin{align}
    &\tau = 0.098 \, z_{\rm rei} - 0.022 \\
    &\tau = 0.001 \, T_0 + 0.065 \, .
\end{align}
Here, $T_0$ is evaluated at the pivot redshift $z_p$ and is given in units of $10^{4}$K. The above dependence on $T_0$ is very weak and for this reason we ignored it in our analysis, simply mapping $\tau \leftrightarrow z_{\rm rei}$. Even though $\tau$ does not have any impact on the SPT analysis, we decided to include in our module the mapping to this CMB parameter in order to have a consistent framework for future analyses that may include CMB data.

Hence, the fully sampled parameter space is made of: 10 parameters for the average flux $\F$ (one for each redshift bin), 2 parameters for each power-law ($T_0$ and $\gamma$), one parameter describing the non-uniform component of the ionizing background ($f_{QSO}$) and 7 cosmological parameters of the $\nu$\LCDM{} model ($A_s, n_s, \oM, \Omega_{b, 0}h^2, \Omega_{\nu, 0}h^2, h, \tau$), for a total of 22 parameters.

\subsection{Galaxy clusters}
The galaxy clusters analysis is performed by following the same procedure as \citet{Bocquet_2019} (henceforth \Boc{}), as we used their - publicly available - \cosmosis{} module. Thus, we only briefly present here the mass proxies used, as well as the results of the analysis, while the reader interested in the details of the methods used can refer to their paper. 

\subsubsection{Mass calibration} \label{sec:mass_calib}
To calibrate the detection significance, $\xi$, the SPT mass proxy, \cite{Bocquet_2019} relied on high quality WL follow-up data.

To account for the impact of noise bias in the measurement of the SZ effect, we characterize the SZ signal with $\zeta$, the detection signal-to-noise ratio (SNR) at the true underlying cluster position and filter scale. This quantity has been shown to be related to the detection significance $\xi$ through a Gaussian distribution with unit scatter and mean $\langle \xi \rangle = \sqrt{\zeta^2+3}$ (\citealt{Vanderlinde10}). We parametrize the $\zeta$-mass relation as
\begin{equation} \label{eq:SZ_scal_rel}
    \begin{split}
    \langle \ln \zeta \rangle = \ln A_{SZ} + B_{SZ} \ln{\left(\frac{M_{500}h_{70}}{4.3\times 10^{14} M_{\odot}}\right)} + C_{SZ} \ln{\left(\frac{E(z)}{E(0.6)}\right)},
    \end{split}
\end{equation}

where $h_{70} = H_0/100/0.7$.
For the WL mass calibration, a bias parameter, $b_{\rm WL}$, is also introduced, so that the relationship between actual cluster mass, $M_{500}$, and its weak-lensing estimate, $M_{\rm WL}$, reads
\begin{equation} \label{eq:WL_mass_proxy}
    \langle \ln M_{\rm WL} \rangle = \ln b_{\rm WL} + \ln M_{500}.
\end{equation}
In the above equations, the redshift and mass dependencies are given in terms of the dimensionless Hubble parameter $E(z)$ and the mass, $M_{500}$, which is defined as the total mass contained within a sphere encompassing a mean overdensity 500 times the critical one. Brackets indicate the expected values for the scaling relations that are assumed to have an intrinsic Gaussian scatter with widths $\sigma_{\ln{\zeta}}$
and $\sigma_{\rm WL}$. We also allow for correlated scatter between the SZ
and WL mass proxies, with all such correlation coefficients being varied along with cosmological parameters.

\Cref{eq:WL_mass_proxy} introduces a bias parameter for the WL mass estimate. This bias accounts for systematic errors in the mass modelling or to systematics in the measurements. The latter are in general related to the instruments and can in principle be different for data obtained with different telescopes/instruments, in our case Magellan/Megacam and HST. In order to take into account such uncertainties, the WL bias can be modelled as
\begin{equation}
\begin{split}
    b_{\rm WL, i} &= b_{\rm WL\, mass,i} \\ &+ \delta_{\rm WL, bias} \Delta b_{\rm WL\, mass\, model, i} \\ &+ \delta_i \Delta b_{\rm syst, i} \\
    &\mathrm{with} \quad i \in \{\rm Megacam, HST\}\,.
\end{split}
\end{equation}
In the above equation $b_{\rm WL\, mass\, model}$ is the mean bias due to mass modelling; $\Delta b_{\rm WL\, mass\, model}$ is the associated uncertainty and $\Delta b_{\rm syst}$ is the systematic measurement uncertainty. Here, $\delta_{\rm WL, bias}$, $\delta_{\rm Megacam}$ and $\delta_{\rm HST}$ are free parameters that we vary in our analysis. At the same time, we model the intrinsic scatter of the mass-observable relations (MOR) due to the fitting of the shear profiles against NFW profiles as
\begin{equation} \label{eq:scatter_WL_shear}
\begin{split}
    \sigma_{\rm WL, i} = \sigma_{\rm intrins, i} + \delta_{\rm WL, scatter} \Delta \sigma_{\rm intrins, i} \\ \mathrm{with} \quad i \in \{\rm Megacam, HST\}.
\end{split}
\end{equation}
Moreover, the intervening large-scale structures can distort the WL shear signal by causing additional deflection of the lensed images and resulting in an error in the cluster mass determination. We model this scatter by following the same approach used above:
\begin{equation}
\begin{split}
    \sigma_{\rm WL, LSS, i} = \sigma_{\rm LSS, i} + \delta_{\rm LSS, i} \Delta \sigma_{\rm LSS, i} \\ \mathrm{with} \quad i \in \{\rm Megacam, HST\},
\end{split}
\end{equation}
where again $\sigma_{\rm intrins}$ and $\sigma_{\rm LSS}$ are the values of the mean scatter associated to the lensing signal of the cluster and of the large-scale structure, respectively; $\Delta \sigma_{\rm intrins}$ and $\Delta \sigma_{\rm LSS}$ are the corresponding errors on such means; $\delta_{\rm WL, scatter}$, $\delta_{\rm LSS, Megacam}$ and $\delta_{\rm LSS, HST}$ are the free parameters, describing the multiplicative telescope noise, which are varied in the chains. The WL modelling framework is described in detail by \cite{Megacam_sample}, while estimates of the modelling parameters introduced in the above equations can be found in Table 1 of \Boc{}.
Through these relations, the observables can be related to the mass of the galaxy clusters which in turn can be linked to the redshift- and mass-dependent HMF. 

The adopted fitting model for the HMF is that of \citet{TinkerHMF},
\begin{equation}
    n(M, z) = \frac{\densm}{M} A\left[ \left(\frac{\sigma_M}{b}\right)^{-a}+1 \right] e^{-c/\sigma_M^2} \abs{\dv{ln\sigma_M}{M}}\,,
\end{equation}
where $\densm$ is the average density of the Universe, $\sigma_M$ is the variance of the density field calculated on spheres that enclose a mass $M$ and $A, a, b, c$ are the fitting parameters which were calibrated against N-body simulations.
The Tinker HMF is strongly dependent on the overdensity $\Delta$ used for determining the mass of the simulated clusters, that in the analsysis by \citet{TinkerHMF} were detected by applying a SO algorithm. In particular, this HMF is calibrated in the range $200 \leq \Delta_{\mathrm{mean}} \leq 3200$ where $\Delta_{\mathrm{mean}}$ is the overdensity with respect to the mean background density. Since we are working with overdensities of 500 with respect to the critical density, the mean spherical overdensity is obtained through $\Delta_{\mathrm{mean}}(z)=500/\Omega_m(z)$.

Based on the simultaneous fit of the mass-observable relations and of the HMF model, the galaxy cluster analysis relies on a multi-observable Poisson log-likelihood. With some algebraic modifications (for details see sec. 3.2 of \Boc{}) this takes the form:

\begin{equation}
\begin{split}
    \ln \mathcal{L}(\theta) = &\sum_i \ln \frac{\mathrm{d}N(\xi, z|\theta)}{\mathrm{d}\xi \mathrm{d} z} \bigg|_{\xi_i, z_i} \\
    &- \int^{\infty}_{z_{\mathrm{cut}}} \mathrm{d}z \int^{\infty}_{\xi_{\mathrm{cut}}} \mathrm{d}\xi \left[\frac{\mathrm{d}N(\xi, z|\theta)}{\mathrm{d}\xi \mathrm{d} z} \right] \\ 
    &+ \sum_j \ln P(g_t|\xi_j, z_j, \theta) \big|_{g_{t_j}}.
\end{split}
\end{equation}
where the first sum runs over all clusters $i$ in the survey and the second one over all clusters $j$ for which follow-up WL tangential shear, $g_t$, measurements are available.
Such form explicitly separates the contribution to the likelihood from the SZ clusters abundance (the first two terms) and from the mass calibration measurements (the third term). 
The two integrals of the second term are truncated by the survey selection function and more precisely with $\xi_{\mathrm{cut}}=5$ and $z_{\mathrm{cut}}=0.25$.
We refer to sec. 3.2 of \Boc{} for details on the definition of the likelihood and how each term is computed.

\section{Results} \label{sec:results}
Using the data sets presented in \cref{sec:data} and encoded in the \cosmosis{} modules presented in \cref{sec:methods}, we derived posteriors on the relevant parameters by carrying out the analyses of the \lymana{} and SPT cluster number counts separately. The exploration of the parameter space has been carried out using \multinest{}\footnote{\url{https://github.com/farhanferoz/MultiNest}} \citep{multinest}, a multi-modal nested sampler (see e.g. \citealt{nested_sampling} for a description of nested sampling) which exploits sophisticated proposal algorithms to efficiently sample multi-modal distributions. 

We present in this section the results of the \lymana{} analysis and of the SPT analysis, at first separately. For these two analysis, we show the best-fit models in \cref{fig:best_Pk} and \cref{fig:best_NC}, respectively. While we already anticipate some discussion on the posteriors obtained with the two analyses in two separate subsections, their contour plots are shown jointly in \cref{fig:triangle_comp}. 

All the constraints on parameters and contour plots shown in this section have been obtained using GetDist\footnote{\url{https://getdist.readthedocs.io/}} (\citealt{getdist}), a Gaussian kernel density estimator that takes the weighted MCMC chains and uses them to produce marginalized histograms to easily visualize the posterior. Unless otherwise stated, the multi-dimensional posterior plots always show the 68 per cent and 95 per cent contours of iso-confidence level.

\subsection{\texorpdfstring{\lymana{}}{lyman} forest} \label{subsec:Lya results}
For our reference \lymana{} analysis we used the same priors on astrophysical parameters as those used by \Vid{}. In particular, Gaussian priors are used on $\F(z)$ for each redshift bin, 
centered on the values obtained in \cite{BOSS_2013} and with a standard deviation of 0.04, such that the results are not dominated by the prior. Wide, non-informative priors are set for the remaining parameters, with the exception of $\gamma$. For this parameter, a flat prior is set over the full redshift range in order to restrict it to physically plausible values. The parameters varied and their priors for the reference case are summarized in \cref{tab:Ly_priors}. Here, and in all the tables henceforth, $\mathcal{U} (min, max)$ denotes the uniform distribution between $min$ and $max$, while the normal distribution with mean $\mu$ and standard deviation $\sigma$ is indicated by $\mathcal{N}(\mu, \sigma)$.

\begin{table}
    \centering
    \begin{tabular}{c c | c c}
        \hline
        \hline
         Parameter & Prior & Parameter & Prior \\
         \hline
         $f_{QSO}$ & $\mathcal{U} (0, 1)$ & $\F(z=3.4)$ & $\mathcal{N}$(0.562, 0.04) \\
         $z_{\rm rei}$ & $\mathcal{U} (7, 15)$ & $\F(z=3.6)$ & $\mathcal{N}$(0.519, 0.04) \\
         $T_0^S$ & $\mathcal{U} (-5, 5)$ & $\F(z=3.8)$ & $\mathcal{N}$(0.467, 0.04) \\
         $T_0^A$ & $\mathcal{U} (0.3, 2)$ & $\F(z=4.0)$ & $\mathcal{N}$(0.419, 0.04) \\
         $\gamma^S$ & $\mathcal{U} (-5, 5)$ & $\F(z=4.2)$ & $\mathcal{N}$(0.364, 0.04) \\
         $\gamma^A$ & $\mathcal{U} (0, 2)$ & $\F(z=4.6)$ & $\mathcal{N}$(0.30, 0.04) \\
         $\F(z=3.0)$ & $\mathcal{N}$ (0.681, 0.04) & $\F(z=5.0)$ & $\mathcal{N}$(0.18, 0.04) \\
         $\F(z=3.2)$ & $\mathcal{N}$(0.625, 0.04) & $\F(z=5.4)$ & $\mathcal{N}$(0.08, 0.04) \\
        \hline
        \hline
    \end{tabular}
    \caption{Priors for the astrophysical parameters used in the \lymana{} analysis. Each of the $\F(z)$ parameters also has a flat prior in the range $(0, 1)$. The prior on $z_{\rm rei}$ reflects the boundaries of the available simulations and that on $T_0^A$ poses a physical limit on the thermal history of the IGM. A tighter prior is set manually on $\gamma$, such that $1<\gamma(z)<1.7$ at each redshift bin.}
    \label{tab:Ly_priors}
\end{table}

We have also set priors on the cosmological parameters and these priors are shared by the two analyses. 
In particular, we have set Gaussian priors on those parameters that are constrained neither by the \lymana{} spectra, nor by the cluster number counts, namely $\ob h^2$, $\onu h^2$ and $h$. Such priors correspond to the marginalized posteriors obtained from CMB data in \citet{Planck18-VI}, but with twice as large standard deviation. As for $\onu h^2$, we used a Gaussian prior centered on the minimum value allowed by oscillation experiments \citep{de_Salas_2018}, with a reasonably large width to allow for deviations from it. On the remaining cosmological parameters, we have set flat uninformative priors which are listed in \cref{tab:cosmo_priors} along with the aforementioned Gaussian priors. 

\begin{table}
    \centering
    \begin{tabular}{c|c c c}
    \hline
    \hline
        Parameter & Prior \\
    \hline 
        $n_s$ & $\mathcal{U} (0.94, 1)$ \\
        $A_s$ & $\mathcal{U} (10^{-10}, 10^{-8})$ \\
        $\oM$ & $\mathcal{U}(0.1, 0.6)$ \\
        $\ob h^2$ & $\mathcal{U}(0.02, 0.024) \times \mathcal{N}(0.02242, 0.00028)$ \\
        $\onu h^2$ & $\mathcal{U}(0, 0.01) \times \mathcal{N}(0.0006, 0.0002)$ \\
        $h$ & $\mathcal{U}(0.55, 0.9) \times \mathcal{N}(0.6766, 0.0084)$ \\
        \hline
        \hline
    \end{tabular}
    \caption{Priors on the cosmological parameters shared by the \lymana{} analysis and the SPT cluster number counts analysis. The Gaussian priors are taken from the results of \citet{Planck18-VI}, by doubling their reported standard deviations. 
    The prior on $\onu h^2$ is centered on the minimum value allowed by neutrino oscillation experiments. The last column lists the values used when fixing the parameters.}
    \label{tab:cosmo_priors}
\end{table}

\begin{figure}
	\includegraphics[width=\columnwidth]{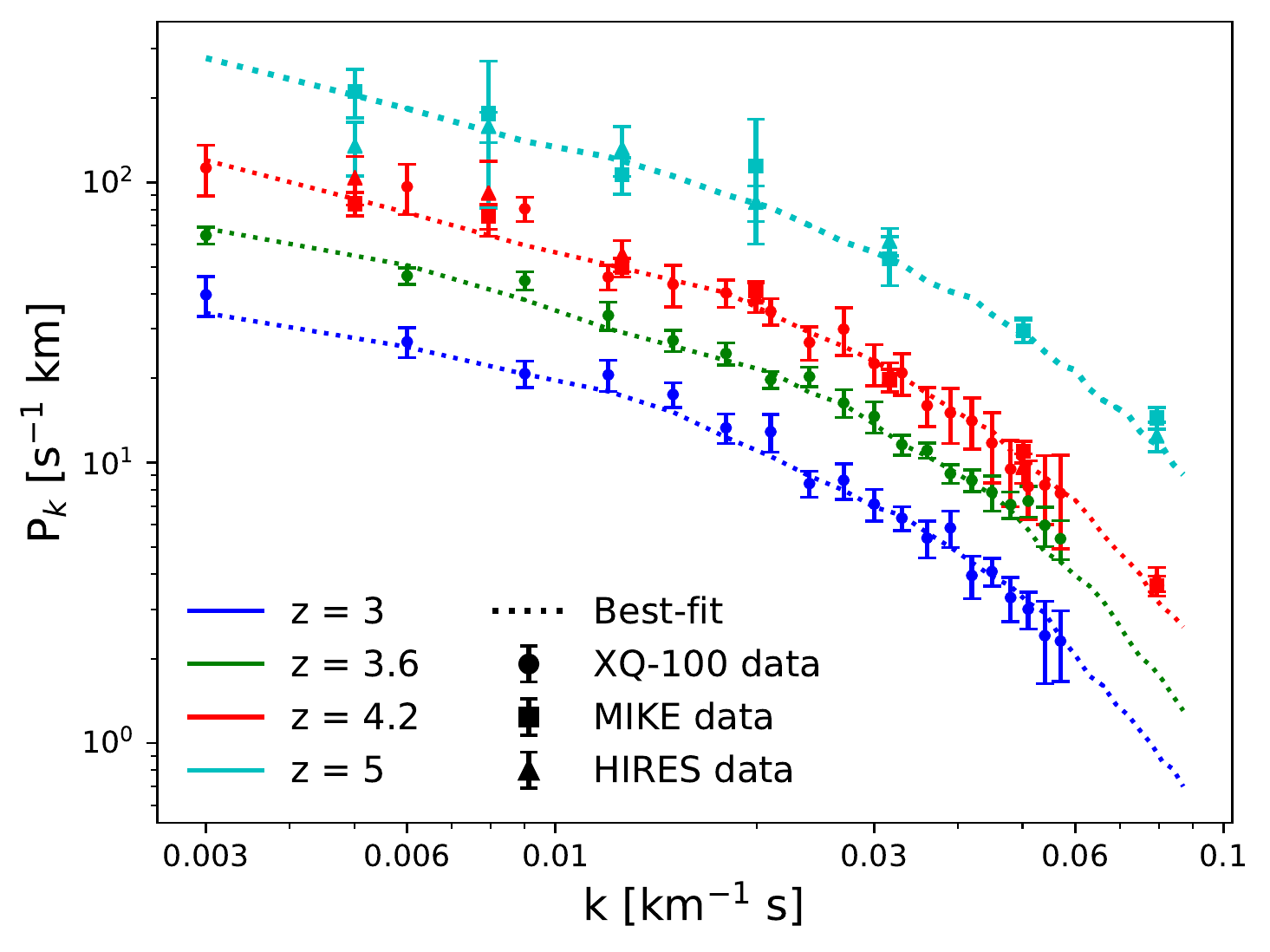}
    \caption{Simulated $P_F(k)$ for the best-fit parameters of our reference \lymana{} analysis for selected redshift bins. Different colors refer to different redshifts, while the shape of the markers indicates which survey each data point belongs to.}
    \label{fig:best_Pk}
\end{figure}

The constraints obtained with such priors are shown as red contours in \cref{fig:triangle_comp} (see below for a discussion on the comparison to the galaxy cluster analysis results). As expected, our \lymana{} data can provide interesting constraints on $\sotto$ and $\neff$. Such parameters are a combination of $A_s$, $n_s$ and $\oM$, which thus exhibit a strong correlation among each other. We have tried using different priors on the cosmological parameters and in particular we confirm that we retrieve the one-to-one mapping between $A_s - \sigma_8$ and $n_s - n_{eff}$ when all the other cosmological parameters are fixed. On the other hand, since we constrain directly $\sotto$, when $A_s$, $n_s$ and $\oM$ are free to vary we see no much degeneracy with any of these parameters.
The best-fit parameters provide a good fit to the data with a $\chi^2 = 181.48$ for 173 degrees of freedom ($d.o.f.$). This is also shown in \cref{fig:best_Pk}, where we plot the simulated flux power spectrum against the data, at different redshifts.

The constraints on $\neff$ show that part of the explored parameter space along that direction extends outside of the range covered by the simulations. In order to obtain prediction for those values of $\neff$ we resort to linear extrapolation. This might introduce an error larger than the one estimated for the interpolation scheme ($< 5\%$) and could indeed produce biased results. However, we do not expect any large impact in the $\sotto$ constraints given that the correlation between these two parameters is present, but not very large. Nonetheless, future work may want to explore this by running hydrodynamical simulations covering a larger range of $\neff$.

As mentioned above, our reference analysis uses a power-law to parametrize the evolution of the IGM temperature at mean density, $T_0$. In order to allow for more freedom in the IGM thermal history, we also ran an analysis with independent $T_0$ values at each redshift bin. We have ran two chains with and without informative priors and in every case we have found consistent contours for the cosmological parameters, showing that those constraints are robust with respect to the redshift evolution of $T_0$.


\subsection{Galaxy clusters}
For our reference analysis of the number counts of SPT clusters, we used the same cosmological priors as in the \lymana{} analysis (see \cref{tab:cosmo_priors}). Along with them, we also varied a set of nuisance parameters which characterize the mass-observable relations:

\begin{equation}
    \begin{split}
    \Theta_{MOR} = \{ &A_{SZ}, B_{SZ}, C_{SZ}, \sigma_{\ln \zeta}, \\ 
    & \rho_{SZ-WL}, \delta_{WL, bias}, \delta_{WL, scatter}, \\
    & \delta_{Megacam}, \delta_{LSS, Megacam}, \delta_{HST}, \delta_{LSS, HST} \},
    \end{split}
\end{equation}
where $\rho_{SZ-WL}$ is the correlation coefficient (i.e. the normalized covariance) between the SZ and WL mass proxies; it can assume values in the range $[-1, +1]$, where $\rho_{SZ-WL} > 0$ $(\rho_{SZ-WL} < 0)$ indicates that the two random variables are correlated (anti-correlated) and $\rho_{SZ-WL} = 0$ that they are totally uncorrelated.

The original work of \Boc{} also included X--ray Chandra measurements that are used along with the WL data to calibrate the SZ mass-observable relation. 
These are measurements of $Y_X=M_g T_X$ - where $M_g$ is the gass mass and $T_X$ is the X--ray temperature of the cluster - a typical quantity used in X--ray cluster cosmology \citep{Kravtsov_2006}.
We decided to not include the $Y_X$ fitting parameters and to restrict our analysis to a calibration through WL data alone. The reason behind this choice is that the inclusion of X-ray data does not improve the constraints on the cosmological parameters and only impacts on the posterior of $\sigma_{\ln \zeta}$, as shown in \Boc{} (see their appendix B). Thus, following their approach, we dropped the X-ray mass calibration and instead set a prior on $\sigma_{\ln \zeta}$ to speed up convergence. Additionally, we also set Gaussian priors on the free parameters $\delta$, appearing in eqs.(4) and (5), so as to account for the Gaussian distribution of the uncertainties in the systematics $\Delta$ of the WL mass modelling. To this purpose, we assume for these $\delta$ parameters Gaussian priors with zero mean and unity standard deviation. All the priors on the relevant parameters are reported in \cref{tab:spt_priors}. Among the reported uniform priors, the one on $\rho_{SZ-WL}$ spans the full range of possible values, while those on the SZ MOR parameters are set to tighter, yet non-informative, ranges based on previous analyses.

\begin{table}
    \centering
    \begin{tabular}{c c}
    \hline
    \hline
       Parameter &Prior\\
    \hline 
        $A_{SZ}$ & $\mathcal{U}(3, 10)$ \\
        $B_{SZ}$ & $\mathcal{U}(1.2, 2)$ \\
        $C_{SZ}$ & $\mathcal{U}(-1, 2)$ \\
        $\sigma_{\ln \zeta}$ & $\mathcal{U}(0.01, 0.5)\times \mathcal{N}(0.13, 0.13)$ \\
        $\rho_{SZ-WL}$ & $\mathcal{U}(-1, 1)$ \\
        $\delta_{LSS, HST}$ & $\mathcal{U}(-3, 3)\times \mathcal{N}(0, 1)$\\
        $\delta_{WL, bias}$ & $\mathcal{U}(-3, 3)\times \mathcal{N}(0, 1)$\\
        $\delta_{Megacam}$ & $\mathcal{U}(-3, 3)\times \mathcal{N}(0, 1)$\\
        $\delta_{HST}$ & $\mathcal{U}(-3, 3)\times \mathcal{N}(0, 1)$\\
        $\delta_{WL, scatter}$ & $\mathcal{U}(-3, 3)\times \mathcal{N}(0, 1)$\\
        $\delta_{LSS, Megacam}$ & $\mathcal{U}(-3, 3)\times \mathcal{N}(0, 1)$\\
        \hline
        \hline
    \end{tabular}
    \caption{Priors on the MOR parameter, following those used in the analysis of \Boc{}.}
    \label{tab:spt_priors}
\end{table}

The marginalized contours obtained from the galaxy clusters analysis are shown in \cref{fig:triangle_comp}. The figure shows that the SPT data does not constrain $n_s$, but as we expected, can constrain well $\oM$ and $\sigma_8$, pointing to lower values of $\sigma_8$ with respect to those obtained in the \lymana{} analysis. Notably, the contours that we obtained for $\sotto$ are not identical to those published in \Boc{}. 
This is due to the fact that in this analysis we use the "neutrino prescription" \citep{Costanzi_2013} for the HMF i.e. we use the cold dark matter + baryons power spectrum for the computation of the Tinker HMF, in opposition to what is done in \Boc{} in which the total matter power spectrum is used instead. In addition to that, we assume 3 degenerate massive, while in \Boc{} the analysis was based on a single massive neutrino. These differences result in the shift in $\sotto$ shown in \cref{fig:all_s8} as pointed out in \cite{Costanzi_2021} (see their note number 9 for further details).

\begin{figure}
    \centering
    \includegraphics[width=\columnwidth]{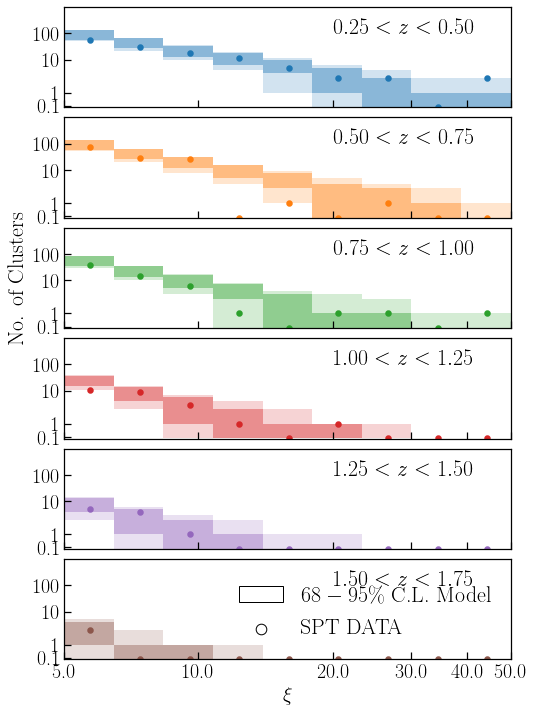}
    \caption{Cluster number counts as a function of the SPT detection significance $\xi$ in 6 different redshift bins. The filled circles represent the SPT cluster number counts, while the shaded areas include the model predictions at the 68 per cent and 95 per cent C.L. The error bars are obtained assuming a Poisson distribution for the predicted counts. The fourth $\xi$ bin at $0.50 < z < 0.75$ shows a strong discrepancy between the observed and the predicted count, raising the value of the C-statistic (see text for details).}
    \label{fig:best_NC}
\end{figure}

To evaluate how well the best-fit model can reproduce the data, we follow the approach used in \Boc{} and evaluate the C-statistics \citep{Cash79} instead of the $\chi ^2$, as the latter is not reliable in our case due to the small number of cluster counts within each redshift and $\xi$ (or mass) bin \citep{Kaastra17}. The $(z,\xi)$ plane sampled by the SPT clusters is divided into 6 evenly spaced $z$-bins between $z=0.25$ and $z=1.75$ and 9 log-spaced $\xi$-bins between $\xi=5$ and $\xi=50$.
With this binning, the expected mean $C_e$ and standard deviation $C_{std}$ calculated from the best-fit simulated number counts  are\footnote{To calculate these values we used the python package {\sc cashstatistic} \url{https://github.com/abmantz/cstat}}:
\begin{equation}
    C_e = 39.0 \qquad C_{std} = 8.3
\end{equation}
Confronting this with the data statistic for our sample,
\begin{equation}
    C_d = 62.4,
\end{equation}
we found a discrepancy between the expected and measured values. We can track down the reason for the bad value of $C_d$ to a particular bin which contains no cluster, despite its expectation value of $\sim 7$ counts. This is shown in \cref{fig:best_NC} (see the fourth $\xi$-bin in the second panel from the top) where we show the counts for each of the redshift bins as a function of $\xi$\footnote{Interestingly, a similar deficit of massive cluster in the redhisft bin $0.5<z<0.65 $ is also observed in the optical DES Y1 redMaPPer catalog, which overlap by $\sim$ 1300 deg$^2$ with the SPT SZ survey footprint (see e.g. Fig. 4 in \citealt{Costanzi_2021}).}. Overall, the agreement is qualitatively good, showing that despite the outlier bin, the model can reproduce the data quite well. 
Removing this single redshift and mass bin, and evaluating the C-statistic again, gives 
\begin{equation}
    C_e = 38.0 \qquad C_{std} = 8.2 \qquad C_d = 48.9\,,
\end{equation}
thus supporting the indication that most of the disagreement between the observed SPT cluster counts and the model-predicted ones is due to this single outlier.

\subsection{Comparing the cosmological constraints}
\begin{figure*}
    \centering
    \includegraphics[width=0.9\textwidth]{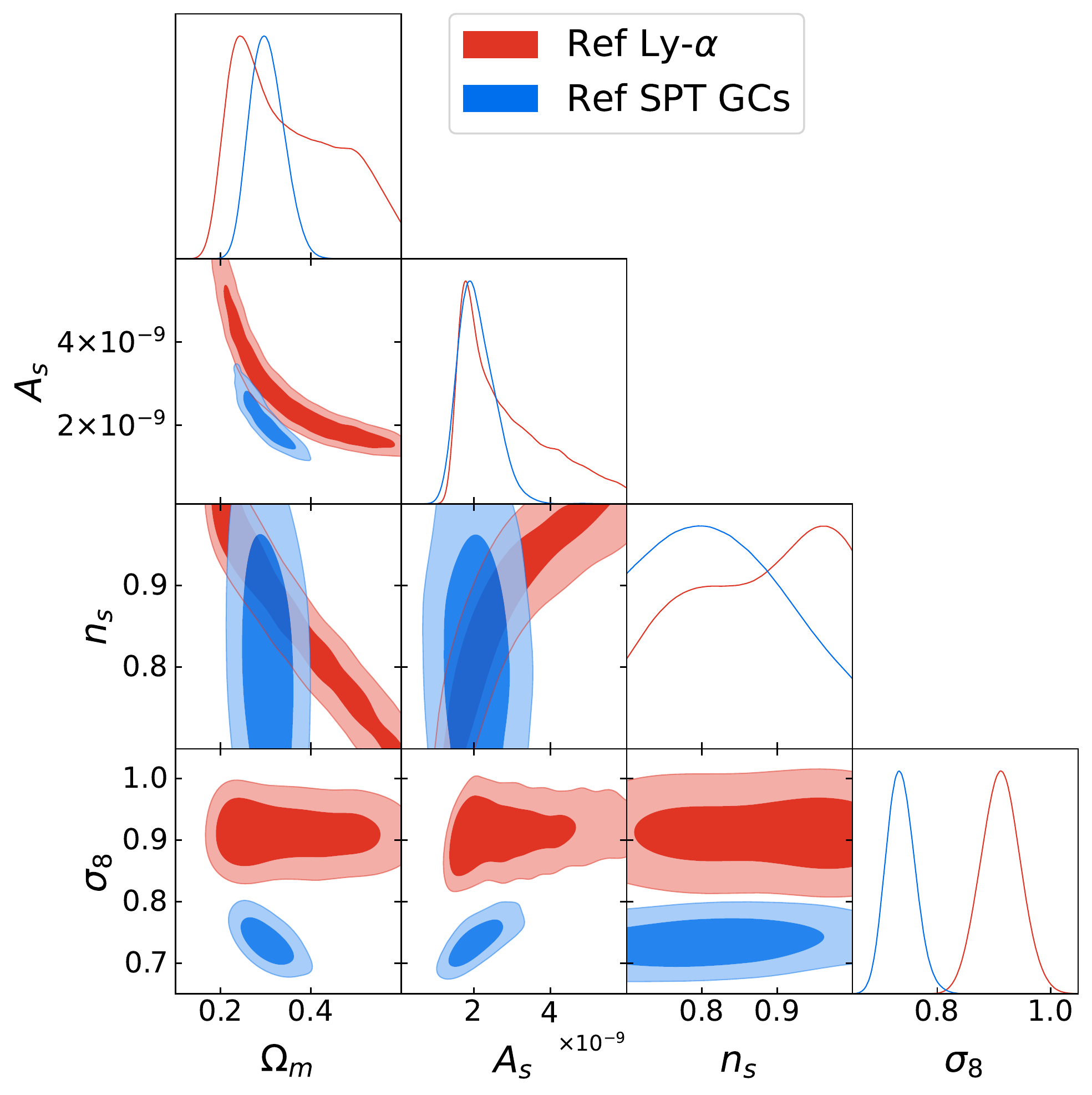}
    \caption{Triangle plot of selected parameters for the chains obtained with the SPT cluster number counts (blue) and the XQ-100 and HIRES/MIKE Lyman-$\alpha$ spectra (red). The marginalized posteriors are plotted only for $\oM$, $A_s$, $n_s$ and $\sigma_8$, because these are the parameters effectively constrained by the two data sets. For the other cosmological parameters, the contours only reflect the prior distributions. The filled and shaded areas represent respectively the 68 per cent and 95 per cent C.L. contours.}
    \label{fig:triangle_comp}
\end{figure*}

We have anticipated in the previous sections that the SPT cluster number counts can effectively constrain $\oM$ and $\sigma_8$, while the HIRES/MIKE and XQ-100 \lymana{} spectra constrain $\sigma_8$ and $n_s$. Therefore, $\sigma_8$ is the parameter which is constrained in independent ways by SPT clusters and by \lymana{} data. For this reason, before  combining the likelihoods of these two analyses, it is fair to compare the two $\sigma_8$ posteriors obtained from the two probes. In fact, we should make sure that we are not combining constraints from independent data which are incompatible at a high significance. In that case, the results of the joint analysis must be handled with great care as they might point to erroneous conclusions, singling out a region of the posterior that is not favoured by any of the data sets separately. In \cref{tab:comp_table} we present the marginalized constraints obtained from the two analyses for the common parameters and we display the relative contours in the triangle plot of \cref{fig:triangle_comp}. 

\begin{table}
    \centering
    \begin{tabular}{c|c c|c c}
    \hline
    \hline
          &\multicolumn{2}{c|}{Galaxy Clusters} &\multicolumn{2}{|c|}{Lyman-$\alpha$} \\
        Parameter & $95\% C.L.$ & Best fit & $95\% C.L.$ & Best fit \\
    \hline
         $\oM$ & $[0.21, 0.40]$ & 0.34 & [0.18, 0.56] & 0.35 \\
         $A_s$ [$10^{-9}$] & $[0.80, 4.11]$ & 1.62 & $[1.32, 5.72]$ & 2.48 \\
         $n_s$ & $[0.7, 1.0]$ & 0.91 & $[0.7, 1.0]$ & 0.84 \\
         $\sigma_8$ & $[0.68, 0.81]$ & 0.74 & $[0.84, 0.98]$ & 0.90 \\
    \hline
    \hline
    \end{tabular}
    \caption{Marginalized constraints at 95 per cent C.L. obtained from the galaxy clusters, the \lymana{} and the combined analysis, when using informative priors on both cosmological and nuisance parameters (see text). The best-fit values are also shown.}
    \label{tab:comp_table}
\end{table}

It is clear from \cref{fig:triangle_comp} that there is a tension on the constraints on $\sigma_8$, as the contours obtained from the two probes do not overlap. This may be due to some undetected systematics or insufficient modelling of the observations. The tension could also be the signature of the failure of the standard \LCDM{} model, which could not be able to provide the connection between the scale- and redshift-ranges over which \lymana{} and SPT clusters probe density perturbations. Clearly, before jumping to this far-reaching conclusion, it is necessary to have a clear understanding of all the possible systematics entering in our analysis. 

First of all, we quantify the degree of disagreement between \lymana{} and SPT clusters constraints by following the approach proposed by \cite{Bocquet_2019} by using the publicly available code released by these authors\footnote{\url{https://github.com/SebastianBocquet/PosteriorAgreement}}. 
The code calculates the probability distribution of the difference, $\delta$, between pair of points randomly drawn from the two 1D posteriors that are being tested (in our case the $\sotto$ posteriors). The p-value that the two posteriors trace the same underlying distribution is then evaluated as
\begin{equation}
    p= \int_{D<D(0)} \mathrm{d}\delta D(\delta),
\end{equation}
where $D$ is the probability distribution of $\delta$ and $D(0)$ is its value for $\delta=0$.
Using this method on the $\sotto$ marginalized posteriors obtained with the \lymana{} and the SPT analyses, we obtain:
\begin{equation}
    p_{\sigma_8} = 0.001,
\end{equation}
which corresponds to a significance of about $3.3\sigma$ under the assumption of Gaussian statistics.
This demonstrates that the $\sigma_8$ tension between the two probes is quite significant. 

Interestingly, we find that despite the tension on $\sigma_8$, the marginalized posteriors on $A_s$ obtained from the two analyses overlap perfectly. This may sound surprising due to the expected correlation between $\sigma_8$ and $A_s$ as both these parameters measure the amplitude of the matter power spectrum. However, $\sigma_8$ is a measurement of such amplitude extrapolated at $z=0$ and on a scale of 8 $h^{-1}$Mpc, a scale at which our probes are sensitive. Conversely, $A_s$ represents the primordial amplitude of the matter fluctuations and is constrained by our data only through an extrapolation at large scales of the constraints obtained on the smaller scales directly probed by \lymana{} spectra and SPT clusters. Hence, the mapping between the two has to take into account the other cosmological parameters (namely $\oM$ and $n_s$), which have a role in defining the shape of the power spectrum and which are constrained neither by the SPT number counts nor by the HIRES/MIKE and XQ-100 \lymana{} spectra. This freedom in the remaining parameters, allows for large overlapping constraints on $A_s$, despite the tension on $\sigma_8$. This can be better understood by looking, for instance, at the $A_s - \oM$ panel in \cref{fig:triangle_comp}, where the contours exhibit a tension at almost $2\sigma$.

Hence, the two data sets show a tension on $\sigma_8$ which can be traced to different preferred $A_s$ values, when imposing a strong prior on $n_s$, e.g. as obtainable from the Planck CMB analyses (e.g. the latest Planck release \citep{Planck18-VI} gives at the 95$\%$ C.L. $n_s = 0.9665 \pm 0.0038$). 


\subsubsection{Inclusion of the DLA correction}

Among all the systematics in the \lymana{} analysis, the one with the largest impact could arise from the incomplete excision of the contamination from damped \lymana{} systems (DLA systems) in the observed quasar spectra. Such strong absorption systems are very rare and should be included by hand in the hydrodynamical simulations when extracting mock quasar spectra from such simulations. To avoid this procedure, the commonly adopted approach is rather to identify and eliminate from the data set those portions of spectra that include strong absorption systems. However, for smaller Hydrogen column densities, e.g. Lyman limits systems (LLS) or subDLAs, it can be quite difficult to identify such systems, thus possibly leading to the introduction of some systematics in the interpretation of the spectra. 

Following \citet{DLA_corr}, we modelled the residual contributions to the flux power spectrum from LLSs and subDLAs, assuming that all the small and large DLAs have been successfully removed during data reduction. Adopting the same notation as in \citet{DLA_corr} (see Eq. 8 in their paper), we refer to $\alpha_{LLS}(z)$ and $\alpha_{subDLA}(z)$ as the relative contribution fractions to the flux power spectrum coming from LLSs and subDLAs respectively. We varied these parameters in our chain in two different ways: {\em (i)} as redshift independent parameters; {\em (ii)} as power-laws of $(1+z)$, varying only the amplitude and fixing the slope to the value obtained by fitting table 1 of \citet{DLA_corr}. We point out that the results in terms of cosmological constraints by adopting either one or the other of these two approaches are virtually identical.

\begin{figure}
	\includegraphics[width=\columnwidth]{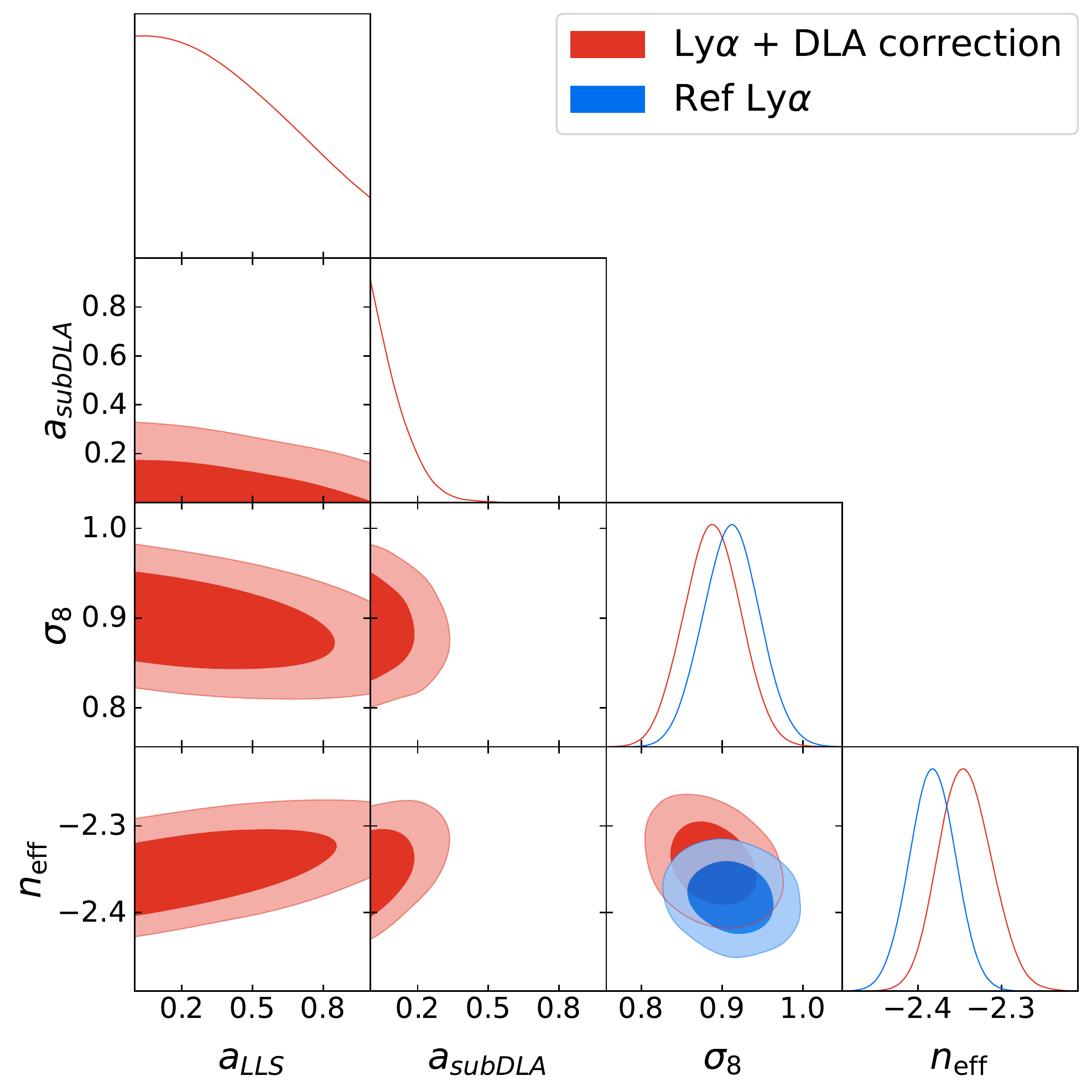}
    \caption{Triangle plot showing the marginalized 1$\sigma$ and 2$\sigma$ contours for $\sotto$ and $\neff$, as well as for the two DLA correction parameters, for the reference \lymana{} chain and the one with the inclusion of the DLA correction parameters. Varying such parameters reduces the tension with the SPT data, but not significantly.}
    \label{fig:DLA_corr}
\end{figure}

The impact of including this correction for strong absorbers (for the case of redshift independent parameters) on the cosmological constraints are shown in \cref{fig:DLA_corr}, where they are compared to the constraints from our reference \lymana{} analysis. Notably, the value preferred for $\alpha_{subDLA}$ is 0, thus suggesting that these systems have been correctly removed from the spectra. Nonetheless, this does not hold for $\alpha_{LLS}$, so that the correction to the flux power spectrum introduced with this parameter shifts the contours for the cosmological parameters towards slightly lower $\sotto$ and higher $\neff$ (see \cref{tab:Lya_DLA_corr_chain_post} for the full table of constraints). This slightly reduces the tension with the SPT constraints, but only down to 2.8$\sigma$.

\subsubsection{Interchanging cosmology}
As a final test, we further investigate by how much we would need to change the astrophysical parameters of the SPT analysis in order to alleviate the tension with the \lymana{} cosmological constraints, and viceversa. Hence, we took the 4D multivariate posteriors obtained for $\oM$, $A_s$, $n_s$ and $\sotto$ in the reference analyses and used them as priors for two new chains interchanging the cosmology.

\paragraph*{\lymana{} analysis with SPT cosmology.}
When assuming in the \lymana{} analysis the cosmological priors from the analysis of the SPT clusters, we relax any priors on the IGM parameters and let them vary freely. In particular, the mean fluxes at each redshift bin are the IGM parameters that have the strongest correlation with $\sotto$ and in fact are the ones that are shifted the most, toward lower values.

The shift in the mean fluxes is not as large as that on $\sotto$, probably because the tension is in this way spread on more parameters and also because the measurement of $\sotto$ is quite robust and the low $\sotto$ value provided by the SPT cosmology prior can only be partially compensated by the other parameters. Confirming this, the $\chi^2$ obtained with this chain is worse ($\Delta[\chi^2] \simeq +4$) with respect to the one obtained with the reference \lymana{} analysis, in spite of an increase in the number of free parameters and thus a decrease in the degree of freedom ($\Delta[\mathrm{d.o.f.}] \simeq-7$).

\begin{figure}
    \centering
    \includegraphics[width=\columnwidth]{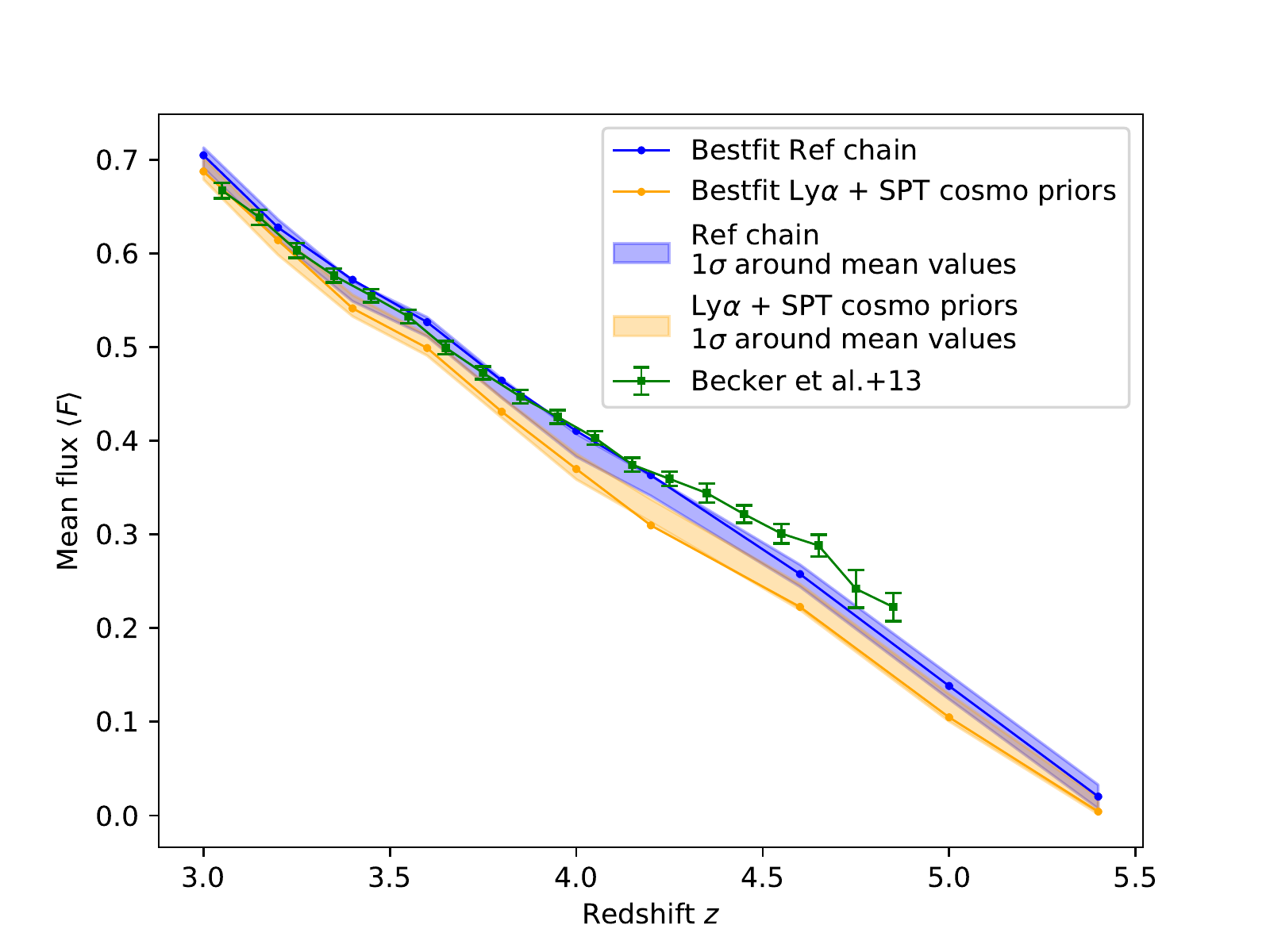}
    \caption{Redshift evolution of the mean fluxes in the bestfit models obtained with the \lymana{} reference analysis and with the analysis with the SPT cosmology. The shaded areas enclose the 1$\sigma$ region around the mean of the posteriors. The points with error bar are independent measurements published in \protect\cite{Becker13}.}
    \label{fig:F(z)_Lya_w_SPT_cosmo}
\end{figure}

Nonetheless, even at this level, the mean fluxes constrained with this analysis exhibit a tension with independent measurements \citep{Becker13}. We show this in \cref{fig:F(z)_Lya_w_SPT_cosmo}, where we confront this measurements with the predicted mean fluxes evolution with redshift for the \lymana{} reference analysis and for the one with the SPT cosmology.
The main uncertainty in determining the mean flux typically resides in the fitting of the continuum of the quasar spectra. Due to large amount of absorption at higher redshifts, this is more of a challenge for MIKE/HIRES data than for XQ-100 or for example BOSS. However, the bias goes in the other direction, such that the mean flux is typically underestimated \citep{Faucher-Giguere08,Wilson21}. Hence, correcting for such a bias would make the tension between the measurements of \cite{Becker13} and the values inferred in the analysis with fixed lower $\sigma_8$ even stronger.


\begin{figure}
    \centering
    \includegraphics[width=\columnwidth]{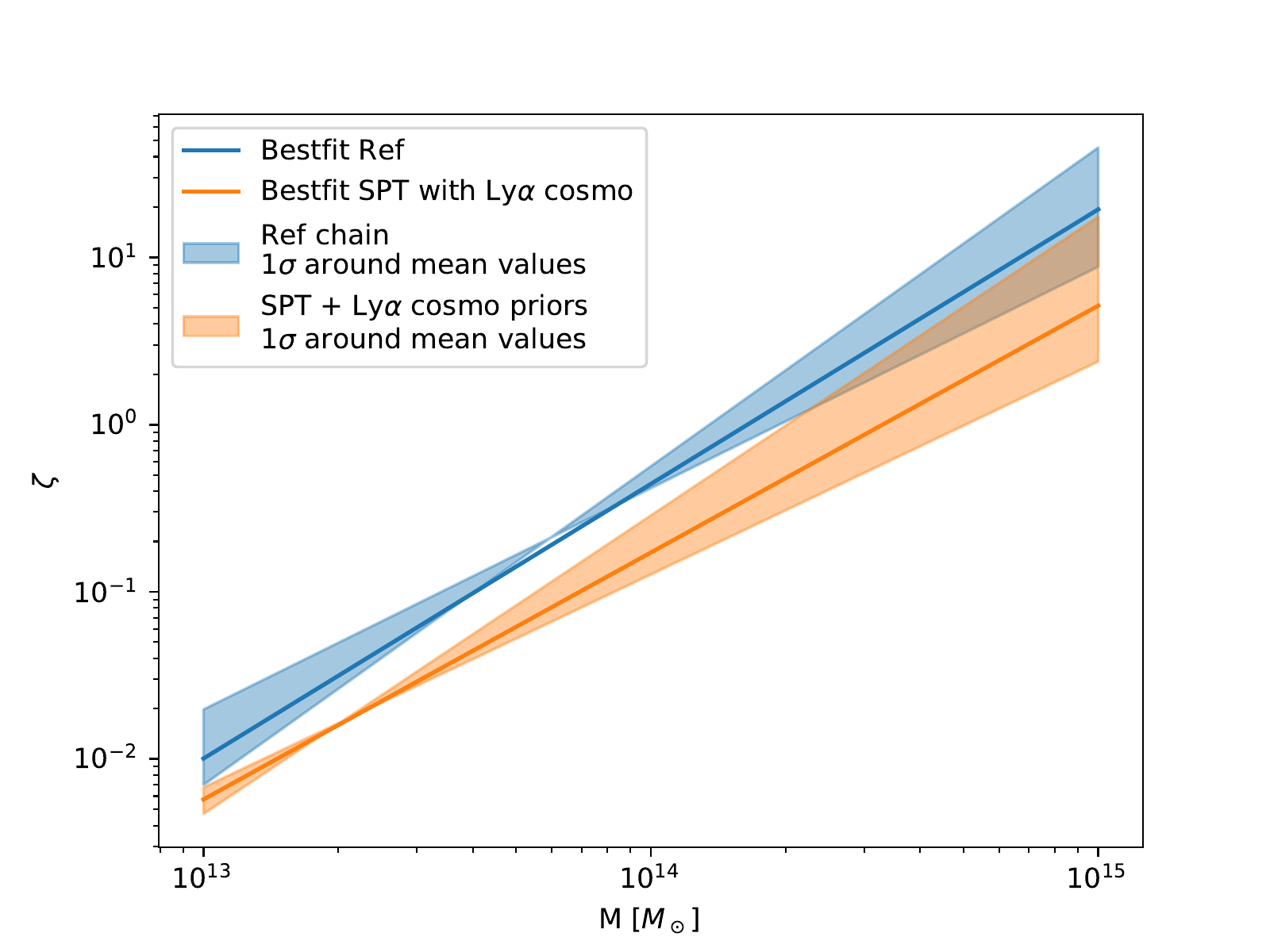}
    \caption{SZ mass scaling relations at the pivot redshift ($z=0.6$) obtained with the SPT reference analysis and with the analysis with the \lymana{} cosmology. The shaded areas enclose the 1$\sigma$ region around the mean of the posteriors.}
    \label{fig:SZ_scaling_rel_w_Lya_cosmo}
\end{figure}

\paragraph*{SPT analysis with \lymana{} cosmology.}
In a similar way, we ran the SPT analysis by using as priors on $\oM$, $A_s$, $n_s$ and $\sotto$ the Gaussian multivariate obtained from the posterior of our reference \lymana{} chain. We did not use the mass calibration adopted in our reference analysis of the SPT cluster counts, and dropped all the priors on the SZ scaling parameters. In this case, the prior on $\sotto$ is informative enough to have a posterior on this parameter that agrees well with the reference \lymana{} one. To obtain such a large $\sotto$ value, the SPT data prefer a very low value for the amplitude of the SZ mass-observable relation, $A_{SZ}$. This in turn means that with the \lymana{} cosmology more massive clusters would correspond to a given SZ signal. We show the SZ scaling relation for the reference SPT analysis and the one with the SPT cosmology in \cref{fig:SZ_scaling_rel_w_Lya_cosmo}.
The disagreement between the amplitude of the reference SZ mass-observable relation used in our analysis and that required to bring SPT cluster counts in agreement with \lymana{} cosmology turns out to be at $\sim 3.4\sigma$ when comparing the marginalized posteriors. This highlights that an exceedingly large change in the SZ mass-observable relation should be invoked to reconcile the cosmological parameters preferred by SPT cluster counts and \lymana{} flux power spectrum.

\section{Summary and conclusions} \label{sec:conclusions}

We presented a comparison of the constraints on cosmological parameters obtained from the analysis of the SPT cluster number counts \citep{Bleem2015} and from the HIRES/MIKE \citep{Viel13} and XQ-100 \citep{XQ100_data} \lymana{} spectra.
These two probes measure the matter power spectrum at complementary scales and over different redshifts ranges. As such, their combination should in principle provide unique constraints on the underlying cosmological model by tracing cosmic growth over more than 90 per cent of cosmic history. 

To judge the feasibility of this multiprobe analysis, we first assess possible tensions between the two data sets by comparing their cosmological posteriors.
To this end, we employed the publicly available likelihood \cosmosis{} module described in \cite{Bocquet_2019} for the SPT analysis and integrated the likelihood scheme presented in \cite{Vid2017} in another \cosmosis{} module for the \lymana{} analysis.

Our main finding in this work is a significant tension
in the determination of the amplitude of the linear matter power spectrum, as described by $\sigma_8$, between the value preferred by the SPT cluster number counts and the \lymana{} flux power spectrum: 
\begin{equation}
    \sigma_8^{\rm GCs} = 0.738^{+0.026}_{-0.040} \qquad \qquad \sigma_8^{\rm Ly\alpha} = 0.911^{+0.034}_{-0.035}\,.
\end{equation}
Using the publicly available Python package described in \cite{Bocquet_2019}, we quantified the degree of disagreement between the two 1D posteriors and found that the two measurements of $\sotto$ differ at $\simeq 3.3 \sigma$. 

In order to further investigate such tension, we performed a list of different tests varying the parameter space and the priors on such parameters:

\begin{itemize}[leftmargin=!,labelindent=5pt,itemindent=-5pt]
    \item 
Using independent z-bins instead of a power-law parametrization for the temperature at mean density $T_0$ in the analysis of the \lymana{} data has a negligible effect on the cosmological posteriors, thus showing the robustness of the $\sotto$ tension with respect to the choice of different methods in our modelling of the IGM.

\item 
Extending the parameter space of the \lymana{} analysis to include a possible residual contribution from subDLA and LLS that may not have been excised from the data,
relaxes the cosmological posteriors and shifts the $\sotto$ posterior to:
\begin{equation}
    \sotto^{DLA} = 0.888 ^{+0.035}_{-0.035}.
\end{equation}
This reduces the tension with the SPT results to $\sim 2.8 \sigma$.
Hence, we conclude that the origin of this tension cannot be due (at least not solely) to the unaccounted effect of DLA systems.


\item 
Analyzing the SPT data adopting the \lymana{} cosmological posteriors as priors,
we found that the shift in $\sotto$ corresponds mainly to a large shift ($\sim 3.4\sigma$) to lower values of the amplitude of the SZ mass-observable relation. Hence, to completely solve the tension the weak lensing mass estimates should be biased low by $\sim 50 \%$.

\item On the other hand, using the $1\sigma$ $\sotto$ marginalized posterior obtained from the SPT analysis as a prior in the \lymana{} analysis is not informative enough to push the contours to such a low $\sigma_8$ value, given the strong constraining power that the \lymana{} data have on such a parameter. Nonetheless, the shift in the $\sotto$ posterior, drives the \lymana{} analysis to values of the mean fluxes that are in clear tension with independent measurements \citep{Becker13}. A bias given by continuum fitting is incompatible with such a shift as it typically points to the opposite direction \citep{Faucher-Giguere08}.

\end{itemize}
\begin{figure}
    \centering
    \includegraphics[width=\columnwidth]{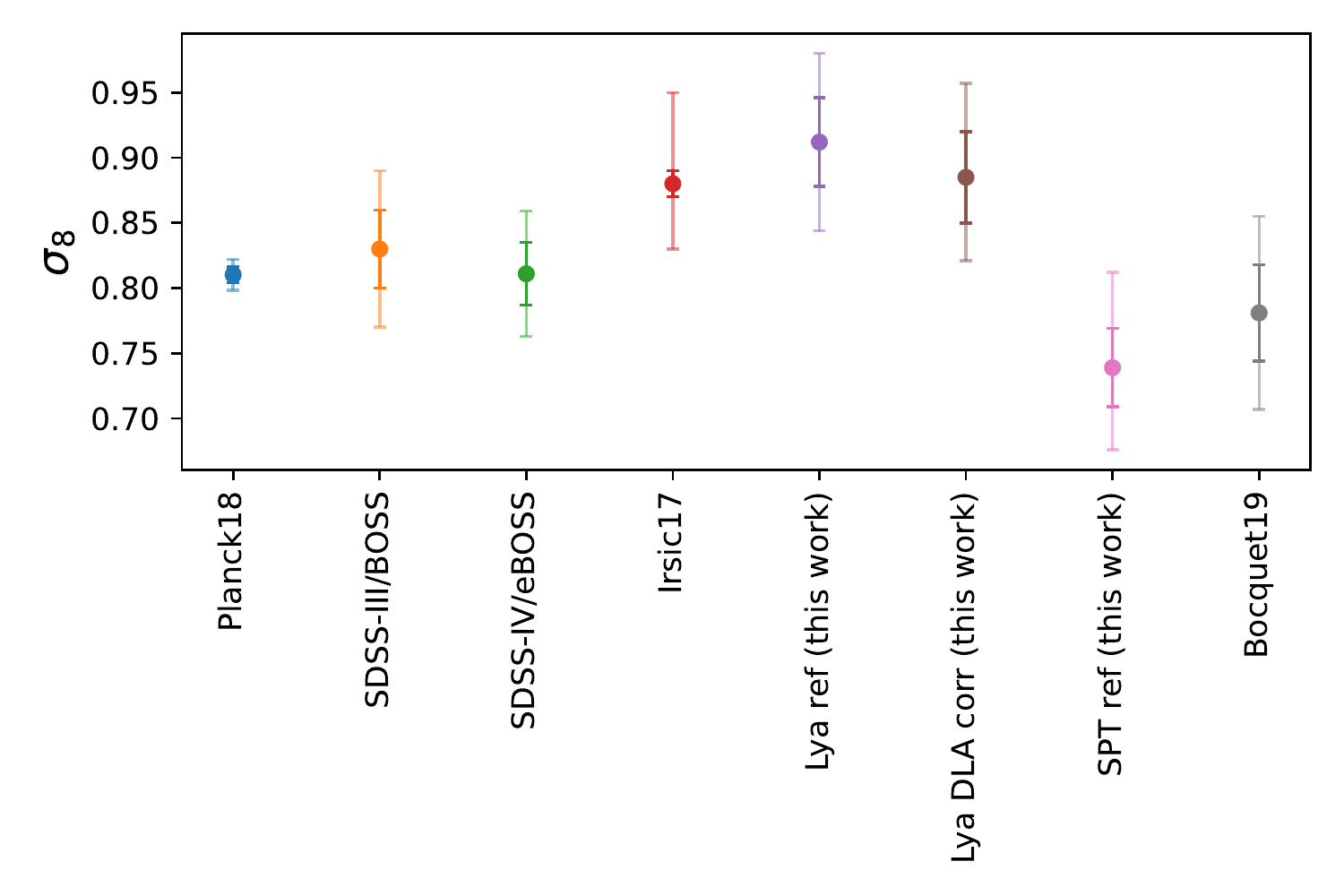}
    \caption{Constraints at 1$\sigma$ and 2$\sigma$ level for $\sotto$ obtained in different analyses. The Planck, SDSS and SPT cluster papers do not report their 2$\sigma$ constraints, hence for those cases we plot twice the 1$\sigma$ ones. On the other hand, for \Vid{} and this works's result, we assume gaussian statistic and show the 68\% and 95\% C.L. as the 1$\sigma$ and 2$\sigma$ contours respectively.}
    \label{fig:all_s8}
\end{figure}


To put things in perspective, we show in \cref{fig:all_s8} the $1\sigma$ and 2$\sigma$ contours obtained using the latest CMB data from Planck \citep{Planck18-VI}, the constraints obtained from the \lymana{} flux power spectrum analyses with SDSS-III \citep{BOSS_2013} and SDSS-IV \citep{eBOSS_2020} data and those obtained in this work, as well as the constraints in the original analyses of \Vid{} and \Boc{} that we have reproduced. 
Note that our \lymana{} results differ from those of \Vid{} because we explored a slightly larger parameter space and because the analysis of \Vid{} had an artefact in the interpolation scheme of cosmological parameters that produced artificially stronger 1$\sigma$ constraints. This however do not affect the 2$\sigma$ contours which are in agreement with our results (see \cref{apx:check_lyman} for details).
Our results for the \lymana{} analysis also show a moderate tension with the BOSS/eBOSS analysis. If our $\sotto$ measurement is just a statistical fluctuation, then the eBOSS analysis could be averaging out such a feature thanks to the larger number of objects included in the survey (while our analysis only uses $\sim100$ spectra, eBOSS has $\sim10000$ spectra for the flux power spectrum measurement). On the other hand, given their different nature in terms of number of objects, resolution and SNR the two analyses may be affected by different systematics (continuum fitting and DLA correction, for example, typically have a stronger effect on larger scales, which are not really sampled by our data; see fig.8 and 9 in \citealt{Karacayli21}) or even physical phenomena.
At the same time, also the SPT constraints are not identical to those published in \Boc{}. The reason for this shift is due to the different prescription used for the halo mass function in the presence of massive neutrinos, as proposed in \cite{Costanzi_2013} and applied to the SPT number counts in \cite{Costanzi_2021}. 
With respect to previous works, the tension in the determination of $\sotto$ is stronger when using the XQ-100 and HIRES/MIKE \lymana{} spectra and the SPT cluster number counts. Moreover, 
the CMB results lie in between the ranges preferred by the analyses performed in this work, also showing a moderate tension of each of them with the Planck data.
It is important to mention in the context of the so-called "$\sotto$ tension" that other LSS probes can constrain this parameter. Some examples are the 3x2 analyses from HSC-Y1 (\cite{Hikage19}, $\sotto \sim 1.05$), KiDS-1000 (\cite{KiDS}, $\sotto \sim 0.76$) and DES-Y3 (\cite{DESY3}, $\sotto \sim 0.70$) or the BAO+RSD eBOSS analysis (\cite{eBOSS2021}, $\sotto \sim 0.81$).

The results of this analysis points towards one or more unmodeled systematics either in the cluster number counts, in the \lymana{} forest spectra or in both probes.
The most likely systematics that could have been underestimated in our analysis are the weak lensing mass calibration for the SPT clusters and the continuum fitting and subDLA contribution for the \lymana{} analysis. However, we have shown in this work that these do not seem to be able to compensate the tension on $\sotto$ unless we assume that such systematics are at a level that can hardly be reconciled with other observational constraints. 

Clearly, a more exciting possibility is that this tension signals a deviation from the standard $\nu$\LCDM{} paradigm, which is always implicitly assumed in our analysis.
It is thus of paramount importance to further investigate the tension. As a first step, extending the analysis to a $\nu w$CDM model will add other degrees of freedom and may (at least partially) relax the tension. Then, more exotic extensions of the standard model could be tested, like for example models with coupled DE-DM, with warm or mixed (cold+warm) DM or scenarios in which the DM is not stable and decays varying its density with time. 

The advent of the ESA's Euclid satellite is expected to provide large samples of galaxy clusters identified in the optical/near-IR photometric survey \citep{Euclid_clusters}. At the same time, the DESI survey will also yield a huge number of QSO spectra over a wide redshift range \citep{DESI}.
This will open up the possibility to carry out similar joint analyses with different and much richer data sets and better understand if this tension is pointing to the need of revising the $\nu$\LCDM{} paradigm or is only given by a limited control of the systematics involved in the cosmological analyses of such two complementary probes. 

\section*{Acknowledgements}
MV, SB, AS are supported by INFN PD51 INDARK grant. AS and MC are supported by the ERC-StG ‘ClustersXCosmo’ grant agreement 716762. AS is also supported by the FARE-MIUR grant 'ClustersXEuclid' R165SBKTMA.
We thank S. Bocquet for useful discussions. VI is supported by the Kavli Foundation.

\section*{Data Availability}

The data underlying this article will be shared on reasonable request to the corresponding author.



\bibliographystyle{mnras}
\bibliography{bibliography} 




\appendix



\section{Consistency check for the \texorpdfstring{\lymana{}}{lyman} module} \label{apx:check_lyman}

\begin{figure*}
    \centering
    \includegraphics[width=0.9\textwidth]{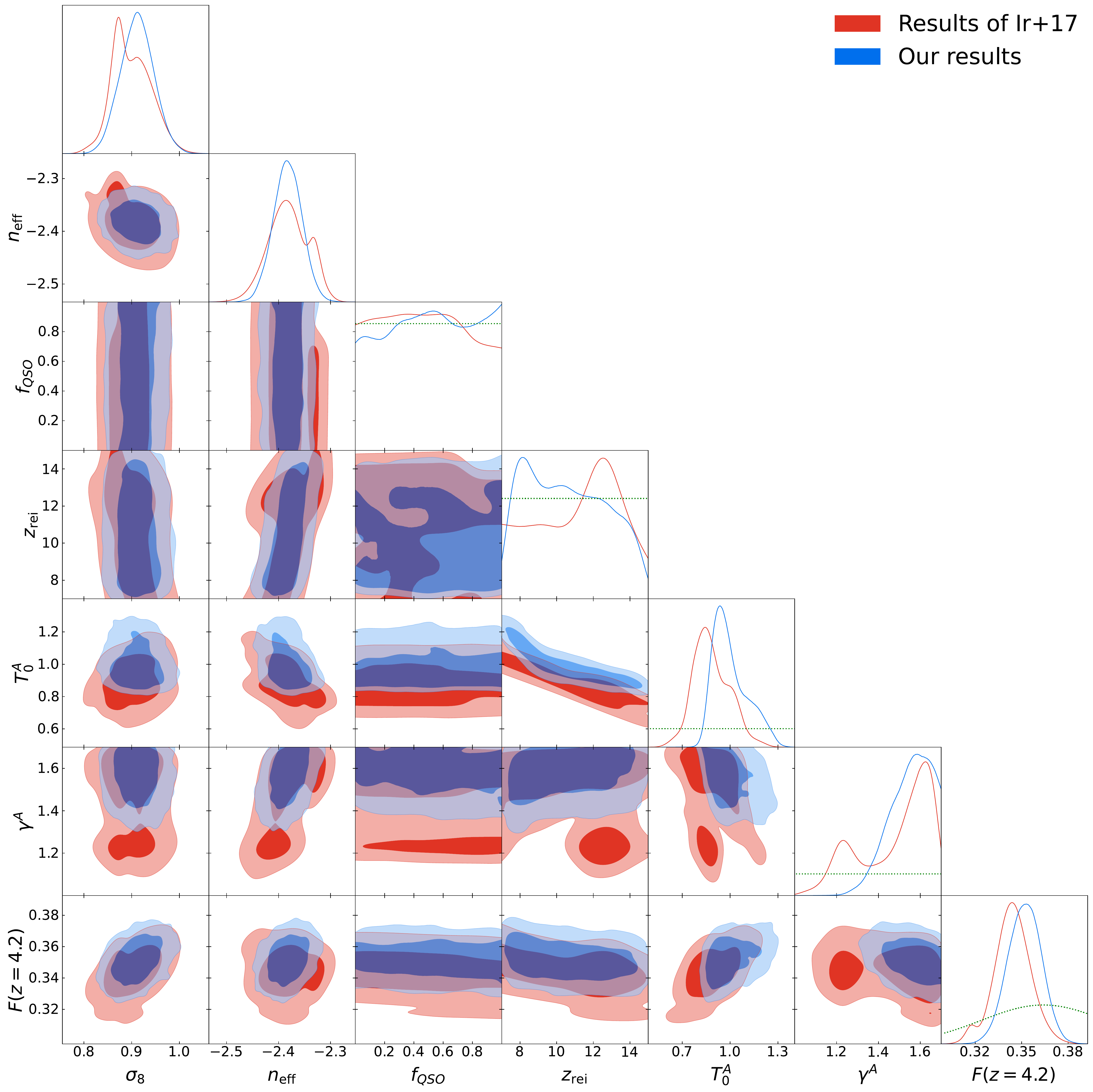}
    \caption{Comparison of the marginalized posteriors obtained with our \cosmosis{} module and those reported in \Vid{} for a selection of parameters. As the redshift pivot of the power-law relations is $z_p=4.2$, $\gamma(z=4.2)$ and $T_0(z=4.2)$ are simply the amplitudes $\gamma^A$ and $T_0^A$ of the power-law parametrizations. The filled and shaded areas represent respectively the 68\% and 95\% C.L. contours. The green dotted lines represent the priors used for the varied parameters. No prior is shown in the $\sotto$ and $\neff$ panels because these are derived parameters.}
    \label{fig:triangle_lyman}
\end{figure*}

In order to validate our \cosmosis{} \lymana{} module, we ran a chain in the same parameter space explored in \Vid{} (i.e. directly varying $\sotto$ and $\neff$ and fixing the other cosmological parameters to Planck-like values).
The results of this analysis are displayed in the triangle plot of \cref{fig:triangle_lyman} for a selection of the varied parameters. We show the contours at 68\% and 95\% C.L. and compare them with the results in \Vid{}. 
The main difference between the two analyses resides in an artefact in the interpolation scheme that has been fixed in this new analysis. In practice, this produced artificially tight constraints at the 68\% confidence level in \Vid{}, but did not affect them at the 95\% confidence level. Moreover, our analysis extends the range of $\F$ values explored in the chain, giving more freedom also in the determination of the other parameters and allowing, for example, the shift seen in $\F($z=4.2$)$.

Another relevant difference between the two analyses resides in the sampling method: while their work is based on a Metropolis-Hastings algorithm, we performed the analysis with a nested sampler. The posteriors have a good qualitative agreement, showing that our pipeline is working properly. Some discrepancy is seen on secondary peaks, due to different starting points or to the different criteria used by the algorithms to determine the number of burn-in samples: in the Metropolis-Hastings algorithm the number of burn-in samples is defined by the user, while in \multinest{} the iteration is terminated when the evaluation of the Bayesian evidence reaches a certain accuracy.

\section{Bestfit and confidence levels tables}

We present in this section the 1 and 2$\sigma$ confidence intervals along with the bestfit values for all the parameters varied in the reference SPT and \lymana{} analysis and in the \lymana{} + DLA correction one. The unconstrained cosmological parameters, i.e. $h$, $\ob h^2$, and $\onu h^2$ have been omitted because the posteriors on those parameters in all the analyses simply reflect the priors.

\begin{table}
    \centering
    \begin{tabular}{c|c|c|c}
    \hline
    \hline
        Parameter & $68\% C.L.$ & $95\% C.L.$ & Best fit \\
    \hline
    $A_{SZ}$ & [5.13, 6.72] & [4.36, 7.71] & 6.38\\
    $B_{SZ}$ & [1.49, 1.63] & [1.41, 1.71] & 1.51\\
    $C_{SZ}$ & [0.55, 1.07] & [0.25, 1.26] & 1.04\\
    $\sigma_{ln \zeta}$ & [0.10, 0.22] & [0.04, 0.29] & 0.17\\
    $\Omega_{m}$ & [0.26, 0.33] & [0.24, 0.38] & 0.34\\
    $A_{s} [10^{-9}]$ & [1.52, 2.44] & [1.24, 3.04] & 1.68\\
    $n_{s}$ & [0.72, 0.89] & [0.70, 1.00] & 0.72\\
    $\sigma_{8}$ & [0.71, 0.76] & [0.69, 0.79] & 0.70\\
    \hline
    \hline
    \end{tabular}
    \caption{Marginalized contours at 68\% and 95\% C.L. as well as bestfit values for the parameters of our reference SPT cluster analysis.}
    \label{tab:SPT_ref_no_mass_calib_chain_post}
\end{table}

\begin{table}
    \centering
    \begin{tabular}{c|c|c|c}
    \hline
    \hline
        Parameter & $68\% C.L.$ & $95\% C.L.$ & Best fit \\
    \hline
    $F(z=3.0)$ & [0.69, 0.71] & [0.68, 0.72] & 0.70\\
    $F(z=3.2)$ & [0.61, 0.64] & [0.60, 0.65] & 0.63\\
    $F(z=3.4)$ & [0.55, 0.57] & [0.54, 0.58] & 0.57\\
    $F(z=3.6)$ & [0.51, 0.53] & [0.50, 0.54] & 0.53\\
    $F(z=3.8)$ & [0.45, 0.47] & [0.44, 0.48] & 0.46\\
    $F(z=4.0)$ & [0.38, 0.41] & [0.37, 0.42] & 0.41\\
    $F(z=4.2)$ & [0.34, 0.36] & [0.33, 0.37] & 0.36\\
    $F(z=4.6)$ & [0.24, 0.27] & [0.23, 0.28] & 0.26\\
    $F(z=5.0)$ & [0.12, 0.15] & [0.11, 0.16] & 0.14\\
    $F(z=5.4)$ & [0.00, 0.03] & [0.00, 0.04] & 0.02\\
    $T_0^A [10^{4} K]$ & [0.86, 1.07] & [0.83, 1.23] & 1.28\\
    $T_0^S$ & [-2.37, -1.15] & [-2.73, -0.45] & -0.93\\
    $\gamma^A$ & [1.51, 1.70] & [1.37, 1.70] & 1.42\\
    $\gamma^S$ & [0.79, 1.53] & [0.35, 1.80] & 1.00\\
    $f_{QSO}$ & [0.00, 1.00] & [0.00, 1.00] & 0.50\\
    $z_{\rm rei}$ & [7.85, 12.57] & [6.87, 14.61] & 7.20\\
    $n_{\rm eff}$ & [-2.41, -2.36] & [-2.44, -2.33] & -2.41\\
    $\Omega_{m}$ & [0.21, 0.45] & [0.18, 0.57] & 0.35\\
    $A_{s} [10^{-9}]$ & [1.37, 3.76] & [0.95, 5.87] & 2.48\\
    $n_{s}$ & [0.82, 1.00] & [0.70, 1.00] & 0.84\\
    $\sigma_{8}$ & [0.88, 0.95] & [0.84, 0.98] & 0.90\\
    \hline
    \hline
    \end{tabular}
    \caption{Marginalized contours at 68\% and 95\% C.L. as well as bestfit values for the parameters of our reference \lymana{} spectra analysis.}
    \label{tab:Lya_ref_chain_post}
\end{table}

\begin{table}
    \centering
    \begin{tabular}{c|c|c|c}
    \hline
    \hline
        Parameter & $68\% C.L.$ & $95\% C.L.$ & Best fit \\
    \hline
    $F(z=3.0)$ & [0.70, 0.72] & [0.69, 0.73] & 0.72\\
    $F(z=3.2)$ & [0.62, 0.65] & [0.61, 0.66] & 0.64\\
    $F(z=3.4)$ & [0.56, 0.58] & [0.55, 0.59] & 0.56\\
    $F(z=3.6)$ & [0.52, 0.54] & [0.51, 0.55] & 0.53\\
    $F(z=3.8)$ & [0.45, 0.47] & [0.44, 0.48] & 0.46\\
    $F(z=4.0)$ & [0.39, 0.41] & [0.38, 0.43] & 0.40\\
    $F(z=4.2)$ & [0.35, 0.37] & [0.34, 0.38] & 0.35\\
    $F(z=4.6)$ & [0.25, 0.27] & [0.24, 0.28] & 0.25\\
    $F(z=5.0)$ & [0.13, 0.15] & [0.11, 0.16] & 0.13\\
    $F(z=5.4)$ & [0.00, 0.03] & [0.00, 0.04] & 0.02\\
    $T_0^A [10^{4} K]$ & [0.86, 1.03] & [0.82, 1.20] & 0.88\\
    $T_0^S$ & [-2.30, -1.11] & [-2.64, -0.36] & -2.53\\
    $\gamma^A$ & [1.56, 1.70] & [1.43, 1.70] & 1.56\\
    $\gamma^S$ & [0.87, 1.63] & [0.41, 1.86] & 1.67\\
    $f_{QSO}$ & [0.00, 1.00] & [0.00, 1.00] & 0.82\\
    $z_{\rm rei}$ & [8.91, 13.65] & [7.32, 14.79] & 14.62\\
    $n_{\rm eff}$ & [-2.38, -2.31] & [-2.40, -2.28] & -2.34\\
    $a_{LLS}$ & [0.00, 0.54] & [0.00, 1.00] & 0.45\\
    $a_{subDLA}$ & [0.00, 0.12] & [0.00, 0.27] & 0.03\\
    $\Omega_{m}$ & [0.26, 0.50] & [0.24, 0.60] & 0.33\\
    $A_{s} [10^{-9}]$ & [1.31, 2.69] & [1.01, 4.05] & 2.34\\
    $n_{s}$ & [0.83, 1.00] & [0.75, 1.00] & 0.91\\
    $\sigma_{8}$ & [0.85, 0.92] & [0.82, 0.96] & 0.89\\
    \hline
    \hline
    \end{tabular}
    \caption{Marginalized contours at 68\% and 95\% C.L. as well as bestfit values for the parameters of our \lymana{} + DLA correction analysis.}
    \label{tab:Lya_DLA_corr_chain_post}
\end{table}





\bsp	
\label{lastpage}
\end{document}